\documentclass[3p]{elsarticle} %3p,draft
 \myfooter[C]{}
\renewcommand{\thispagestyle}[1]{}

\makeatother

\date{}
\def\texpsfig#1#2#3{\vbox{\kern #3\hbox{\includegraphics{#1}\kern #2}}\typeout{(#1)}}

\usepackage{url,hyperref}
\hypersetup{
    colorlinks=true,
    linkcolor=blue,
    filecolor=magenta,      
    urlcolor=cyan,
    pdfpagemode=FullScreen,
    }
\usepackage{bbm}
\usepackage{eurosym}                           % Symbol for Euro
\usepackage[latin1]{inputenc}                  % Input encoding for Western Europe
\usepackage{url}                               % Display URLs in typewriter font
\usepackage{longtable}                         % Tables running of several pages
\usepackage{array}                             % more table options
\usepackage{graphicx,color}
\usepackage{amsthm}
\usepackage{amsbsy}

\newcommand*\bff{\ensuremath{\boldsymbol}}
\usepackage{amssymb}
\usepackage{amsmath}
\usepackage{enumerate}
\usepackage{graphicx,color}
\usepackage{epsfig}
\usepackage[ruled,vlined]{algorithm2e}
\usepackage{multirow,bigdelim}
\usepackage[table]{xcolor}% http://ctan.org/pkg/xcolor
\usepackage{natbib}
\theoremstyle{plain}
\newtheorem{thm}{Theorem}[section]
\newtheorem{cor}{Corollary}[section]

\newtheorem{rem}{Remark}
\theoremstyle{remark}

\theoremstyle{plain}
\newtheorem{lem}[thm]{Lemma}

\theoremstyle{definition}

\newcommand{\e}{{\rm e}}        % "e" number
\def\R{\mathbb{ R}}             % Real number
\def\vix{\mathrm{vix}}             % Real number
\def\VIX{\mathrm{VIX}}             % Real number
\def\E{\mathbb{ E}}             % Expectation
\def\Q{\mathbb{ Q}}             % Measure Q

\def\P{\mathbb{ P}}             % Measure P
\def\F{\mathcal{F}}             % Filtration
\renewcommand{\d}{{\rm d}}      % straight "d" in in integration and ODEs, \int_a^b f(x)\d x
\def\dW{{\rm d}W}               % dW in SDEs- Brownian noise.
\def\dt{{\rm d}t}
\def\ds{{\rm d}s}
\def\du{{\rm d}u}

\def\dx{{\rm d}x}

\def\T{{\rm T}}
\def\1{{\mathbbm{1}}}            % Indicator function

\theoremstyle{plain}% default

\newtheorem{theorem}{Theorem}[section]

\usepackage[margin=1cm]{caption}
\numberwithin{equation}{section}	     %Equation numbering per section

\title{ \raggedright On Randomization of Affine Diffusion Processes\\ with Application to Pricing of Options on VIX and S\&P 500 }

\begin{document}

\author[1,2]{LECH A. GRZELAK\corref{cor1}}
\ead{L.A.Grzelak@uu.nl}
\cortext[cor1]{Corresponding author at Mathematical Institute, Utrecht University, Utrecht, the Netherlands.}
\address[1]{Mathematical Institute, Utrecht University, Utrecht, the Netherlands}
\address[2]{Financial Engineering, Rabobank, Utrecht, the Netherlands}
%{\raggedright ver 1. \today}
\begin{abstract}
The class of Affine (Jump) Diffusion~\citep{Duffie:2000} (AD) has, due to its closed form characteristic function (ChF), gained tremendous popularity among practitioners and researchers. However, there is clear evidence that a linearity constraint is insufficient for precise and consistent option pricing. Any non-affine model must pass the strict requirement of quick calibration- which is often challenging. We focus here on Randomized AD (RAnD) models, i.e., we allow for exogenous stochasticity of the model parameters. Randomization of a pricing model occurs outside the affine model and, therefore, forms a generalization that relaxes the affinity constraints. The method is generic and can apply to any model parameter. It relies on the existence of moments of the so-called randomizer- a random variable for the stochastic parameter. The RAnD model allows flexibility while benefiting from fast calibration and well-established, large-step Monte Carlo simulation, often available for AD processes. The article will discuss theoretical and practical aspects of the RAnD method, like derivations of the corresponding ChF, simulation, and computations of sensitivities. We will also illustrate the advantages of the randomized stochastic volatility models in the consistent pricing of options on the S\&P 500 and VIX.
\noindent 
\end{abstract}

%(RAnD~\footnote{Eng.: ``{\it When awesome and cool just aren't enough to describe something.}'' but also {\it ``a former unit of absorbed ionizing radiation dose equivalent to an energy absorption per unit mass of 0.01 joule per kilogram of irradiated material.''}})

\maketitle

{\let\thefootnote\relax\footnotetext{The views expressed in this paper are the personal views of the authors and do not necessarily reflect the views or policies of their current or past employers. The authors have no competing interests.}}

%%%%%%%%%%%%%%%%%%%%%%%%%%%%%%%%%%%%%%%%%%%%%%%%%%%%%%%%%%%%
\section{Introduction}
%%%%%%%%%%%%%%%%%%%%%%%%%%%%%%%%%%%%%%%%%%%%%%%%%%%%%%%%%%%
The class of Affine Jump Diffusion, which we abbreviate by AD, asset price models constitutes a family of processes that, under certain linearity conditions, allow for a semi-closed form characteristic function (ChF)~\cite{Duffie:2000}. Once the ChF is derived, it can be used, via Fourier inversion, for pricing certain financial derivatives, thus facilitating fast asset price model calibration. Over many years, various affine models have been studied, leading to significant contributions like efficient large-step Monte Carlo simulations, PDE discretizations, quick calibrations or even analytical, perturbation-based approximations for implied volatilities. These accomplishments popularized the class of AD processes. 

Although the high speed of ``regular'' option pricing is a necessary condition that allows for calibration, it is insufficient for a model to be used for exotic derivative pricing. The main limitation of classical dynamic stochastic volatility models driven solely by possibly correlated Brownian motion is the limited flexibility for short maturities, i.e., the slope of the skew is insufficient to fit the market data. By adding jumps, this effect can, to some extent, be mitigated; however, it is often unsatisfactory, especially in a volatile market.  The class of affine models is inherently inflexible, with limited possibilities for model extensions. In particular, with the rise in popularity of products such as options on volatility, it is evident that the standard affine, stochastic volatility models are insufficient for accurate and consistent option pricing. This observation has initiated a {\it gold rush} to discover new classes of models that would address this pricing problem. 

One possible route to extend the existing affine models was suggested by P. Carr and L. Wu in~\cite{carr2007stochastic}, where the problem of insufficient skew was reported, and the remedy of the form of parameter randomization was suggested:  {\it ``Starting from the jump-diffusion stochastic volatility model of Bates~\cite{Bates:1996}, it would be tempting to try to capture stochastic skewness by randomizing the mean jump size parameter or the correlation parameter (...) However, randomizing either parameter is not amenable to analytic solution techniques that greatly aid econometric estimation.''} Although the authors considered randomization as a technique to incorporate more flexibility into a stochastic model, thus a way to improve the calibration quality, the concept of randomizing is more fundamental, i.e., it represents the incorporation of the uncertainty of potentially hidden states that are not adequately captured by deterministic parameters.  In this article, we focus on the randomization of the class of affine models and allow for additional flexibility by letting the model parameters be stochastic. In particular, we will randomize the parameters of the Bates model and examine their impact on the shapes of implied volatilities. 

Attempts to randomize individual model parameters are already known in the literature. Starting from the simplest model, the Black-Scholes model with the randomized volatility coefficient- based on a discrete distribution- was presented in~\cite{brigo2002lognormal}. An extension to continuous distributions was analyzed in~\cite{jacquier2015black}. In their work, the authors discussed various stochastic variants for the volatility and their level of {\it explosiveness}- this led to several valuable expansions for short maturity options. In~\citep{jacquier2019randomized,mechkov2015hot}, for example, it was shown that randomization of the initial variance of the Heston model significantly increased the steepness of the implied volatility, primarily when the randomizer is defined in an unbounded domain. This idea was further studied in~\citep{pages2020stationary} where the term {\it stationary}, to describe random initial parameter, was used. In~\citep{pages2015recursive} the concept of Recursive Marginal Quantization for one-dimensional diffusion discretization schemes was discussed, with the extension to higher order schemes in~\citep{rudd2017fast} and in~\cite{callegaro2019quantization} a method to discretize a stochastic process via quantization optimally was given.  Another randomization attempt was presented in~\citep{fouque2018heston}, where the volatility-of-volatility parameter in the Heston model was assumed to be stochastic, and approximations, based on perturbation, were derived. In the case of a randomizer with a finite number of states, one may think of a regime-switching model~\cite{goutte2017regime}. Such a setup was discussed~\citep{SircarRegimeSwitching}, where the hidden regime continuous-time Markov chain was considered. 

Customarily, an extension with a stochastic parameter would require that the model satisfies the linearity conditions- this, however, does not need to be the case under the framework proposed here. Instead, we only require the affinity to hold for a particular realization of a stochastic parameter. We build an external layer over the class of AD models and allow the parameter to be stochastic in that layer. In a standard approach, this would imply an infinite number of affine models, making this method impractical. The solution to this problem is at the heart of this article, in which we develop a method for constructing an appropriate ChF based on only a few, {\it critical} parameter realizations. Selecting these points is based on a quadrature rule based on the moments of the stochastic parameter. The procedure of building a randomized ChF, we call the RAnD method, which corresponds to ``Randomized Affine-Diffusion''. 

The presented structure is generic, and we believe it may be applied in various asset classes. To highlight its flexibility in modern finance, we apply this technique to the challenging problem of consistent pricing options on S\&P 500 and VIX. This pricing problem is not new, and many innovative solutions have already been proposed.

The major challenge regarding this pricing task is to find a stochastic model that can sufficiently explain both the negative skew for equity options on the S\&P 500 index and the positive skew for options on the VIX. In addition, different regime settings for implied volatility cause a calibration dilemma where the volatility of the volatility parameter under the Heston model is either too large for VIX or too small for S\&P 500 (or both). The first attempts to address this pricing problem aimed to introduce additional degrees of freedom in multidimensional models. In~\cite{sepp2008pricing}, for example, the variance process in the Bates model was enriched with an additional jump process. In~\citep{baldeaux2014consistent}, the so-called 3/2 process and inclusion of jumps for the stock process could capture the observed upward skew for VIX, although the model was unable to calibrate to both derivatives simultaneously. More recently, in~\citep{kokholm2015joint}, another attempt to include jumps in both returns and volatility added to stochastic volatility was discussed. The authors concluded that more flexibility, especially with a relaxed Feller condition, is necessary to provide a sufficient fit. A promising extension through the Hawkes process for the jumps was discussed~\cite{jing2021consistent}- however, additional degrees of freedom are insufficient for calibration purposes. A breakthrough in volatility modelling came with the so-called ``{\it rough volatility}'' models. In~\citep{jacquier2018vix}, the asymptotic expansions of the rough Bergomi~\cite{bayer2016pricing} model to jointly describe the VIX and the S\&P 500 were presented. It was further extended in~\cite{horvath2020volatility} with the application of modulated Volterra processes. Recently, in~\citep{gatheral2020quadratic}, Gatheral et al. have introduced and successfully calibrated the so-called {\it quadratic rough Heston}. Another angle to address this intriguing pricing problem was proposed in~\cite{guyon2020joint}, where a discrete-time dispersion-constrained martingale transport problem was applied. 

The references enumerated above unquestionably address the problem of consistent options pricing on an index and its corresponding volatility. However, in our view, this is insufficient to claim victory- many of the proposed solutions only work for short maturities or lack effective numerical methods to be perceived as practical. We will show that by randomizing AD models, one can benefit from well-established numerical techniques and the ability to price long-term options while consistently pricing volatility products.

The article is organized as follows: in Section~\ref{sec:RAnD}, we start with the definition of the class of AD and give an integral form for the randomized ChF. In Section~\ref{sec:ChFRAnD}, the RAnD method is introduced, and extensions for bivariate and multivariate parameters cases are presented. Section~\ref{sec:RAnD_BS} presents an illustrative example of the randomization of the Black-Scholes model. We analyze here the computational aspects of the method and the evolution of implied volatilities. Section~\ref{sec:Bates} focuses on the randomized Bates model; here, we start with the analysis of randomization and its impact on the implied volatilities; also, in this section, we address what was mentioned earlier in the introduction- the randomization of jump sizes. Further, in Section~\ref{sec:vix_Bates}, we provide the pricing details for VIX options, and in Section~\ref{sec:calibrationVIX}, we use derived formulae to perform simultaneous model calibration to S\&P and VIX options. Discussion about hedging under the RAnD framework is covered in Section~\ref{sec:hedging} and in Sections~\ref{sec:convergence} and~\ref{sec:piece_wise}, the details regarding Monte Carlo simulation with possible improvements are given. Section~\ref{sec:conclusion} concludes.

%%%%%%%%%%%%%%%%%%%%%%%%%%%%%%%%%%%%%%%%%%%%%%%%%%%%%%%%%%%%
\section{Affine (Jump) Diffusion processes and randomization}
\label{sec:RAnD}
%%%%%%%%%%%%%%%%%%%%%%%%%%%%%%%%%%%%%%%%%%%%%%%%%%%%%%%%%%%
The class of stochastic AD processes for the asset dynamics
refers to a fixed probability space $\left(\Omega,\F(t),\Q\right)$
and a Markovian $n$-dimensional affine process
${\bf{X}}(t)=[X_1(t),\dots,X_n(t)]^\T$ in some space $\R\subset
\R^n$. 

The stochastic model of interest can be expressed by the
following stochastic differential form:
\begin{equation}
	\d{\bf{X}}(t)=\bar{\boldsymbol\mu}(t,{\bf{X}}(t))\dt+\bar{\boldsymbol\sigma}(t,{\bf{X}}(t))\d{\widetilde{\bf{W}}}(t)+{\bf J}(t)^\T\d {\bf X}_{\mathcal{P}}(t),
\label{ajd:ch1}
\end{equation}
where ${\widetilde{\bf{W}}}(t)$ is an $\F(t)$-standard column vector of
{\it independent} Brownian motions in $\R^n$,
$\bar{\boldsymbol\mu}{\left(t,{\bf{X}}(t)\right)}:\R\rightarrow\R^n$,
$\bar{\boldsymbol\sigma}\left(t,{\bf{X}}(t)\right): \R\rightarrow\R^{n\times n}$, and
${\bf X}_{\mathcal{P}}(t)\in \R^n$ is a vector of orthogonal Poisson
processes, characterized by an intensity vector
$\bar{\boldsymbol\xi}(t,{\bf X}(t))\in \R^n$. 

${\bf J}(t)\in \R^{n}$ is a vector governing the amplitudes of the jumps and is assumed
to be a matrix of correlated random variables, that are independent
of the state vector ${\bf X}(t)$ and of ${\bf X}_{\mathcal{P}}$(t).

Additionally, we consider an orthogonal vector $\Theta=[{\vartheta}_1,\dots,{\vartheta}_n]^\T$, $n\in\mathbb{N}$, where each ${\vartheta}_i$ is an independent, time-invariant, random variable~\footnote{We consider here $n\in\mathbb{N}$ stochastic parameters, this is however not a necessary constraint.}. A realization of ${\vartheta_i}$ we indicate by $\theta_{i}$, $\vartheta(\omega)=\theta$, {\it per se} the realization for ${\Theta}$ is indicated by $\theta=[\theta_{1},\dots,\theta_n]^\T$.

For processes in the AD class, it is required that for a particular parameter realization, $\theta$, drift $\bar{\boldsymbol\mu}(t,{\bf{X}}(t))$, interest rate component $\bar{\boldsymbol r}(t,{{\bf{X}}(t)})$, and
the covariance matrix $\bar{\boldsymbol\sigma}(t,{\bf X}(t))\bar{\boldsymbol\sigma}(t,{\bf X}(t))^\T$ are of {\em
the affine form}, i.e.
\begin{eqnarray}
\label{1_mu}
	\bar{\boldsymbol\mu}(t,{\bf{X}}(t))&=&a_0(\theta)+a_1(\theta){\bf{X}}(t), \text{ for any } (a_0(\theta),a_1(\theta))\in \R^n \times \R^{n \times n },\\
	\label{1_r} \bar{\boldsymbol r}(t,{\bf{X}}(t))&=&r_0(\theta)+r_1(\theta)^\T{\bf{X}}(t), \text{ for }
(r_0(\theta),r_1(\theta))\in \R \times \R^n,\\
	(\bar{\boldsymbol\sigma}(t,{\bf{X}}(t))
	\bar{\boldsymbol\sigma}(t,{\bf{X}}(t))^\T)_{i,j}&=&(c_0(\theta))_{i,j}+(c_1(\theta))^\T_{i,j}{\bf{X}}_j(t),
\label{1_sigma}\\
\label{1_lambda} \bar\xi(t,{\bf X}(t))&=&l_0(\theta)+l_1(\theta){\bf X}(t), \text{ with
} (l_0(\theta),l_1(\theta))\in\R^n\times \R^{n},
	\label{xifunc}
\end{eqnarray}
with $(c_0(\theta),c_1(\theta))\in \R^{n \times n}\times\R^{n\times n \times n}$, 
and where in (\ref{1_sigma}) it is meant that each element in the matrix $\bar{\boldsymbol\sigma}(t,{\bf{X}}(t)) \bar{\boldsymbol\sigma}(t,{\bf{X}}(t))^\T$ should be affine, as well as each vector element in the drift and interest rate vectors. 

%%%%%%%%%%%%%%%%%%%%%%%%%%%%%%%%%%%%%%%%%%%%%%%%%%%%%%%%%%%%
%\subsection{Randomized Characteristic Function}
\subsection{The RAnD method: details on construction of the characteristic function}
\label{sec:ChFRAnD}
%%%%%%%%%%%%%%%%%%%%%%%%%%%%%%%%%%%%%%%%%%%%%%%%%%%%%%%%%%%
For a given realization of $\Theta$, $\theta$, we consider ${\bf X}_\theta(t):={\bf X}(t)|{\bf \Theta}=\theta$, ${\bf J}_\theta(t):={\bf J}(t)|{\bf \Theta}=\theta$.
It is shown~\citep{Duffie:2000} that in this class, for a state vector ${\bf{X}}_\theta(t)$, the
discounted characteristic function is also of the following form:
\begin{equation}
\label{eqn:ChFCond}
    \phi_{{\bf X}_\theta}({\bf{u}};t,T)=\E_t\left[\e^{-\int_t^T r(s)\ds+i{\bf u}^\T{\bf{X}}_\theta(T)}\right]=\e^{\bar A({\bf u};\tau,\theta)+{\bf{\bar B^\T}}({\bf u};\tau,\theta){\bf{X}}_\theta(t)},
\end{equation}
with the expectation under risk-neutral measure $\Q$ for $\tau=T-t.$ 

The
coefficients $\bar{A}:=\bar A({\bf u};\tau,\theta)$ and
$\bar{\bf B}:={\bf{\bar B}}^\T({\bf u};\tau,\theta)$ have to satisfy
the following complex-valued {\em Riccati} ODEs, see
the work by Duffie-Pan-Singleton~\citep{Duffie:2000}:
\begin{equation}
\label{duffie1a} 
\begin{aligned}
\frac{\d \bar A}{\d\tau}&=-r_0(\theta)+{\bf{\bar B}}^\T a_0(\theta)
+\frac{1}{2}{\bf{\bar B}}^\T c_0(\theta){\bf{\bar B}}+l_0^\T\E\left[\e^{{\bf J}_\theta(\tau){\bf \bar B}}-1\right],\\
\frac{\d\bf{\bar B}}{\d\tau}&=-r_1(\theta)+a_1(\theta)^\T{\bf{\bar B}}+\frac{1}{2}{\bf{\bar B}}^\T
c_1(\theta){\bf{\bar B}}+l_1(\theta)^\T\E\left[\e^{{\bf J}_\theta(\tau){\bf \bar B}}-1\right],
\end{aligned}
\end{equation}
where the expectation, $\E[\cdot]$ in~(\ref{duffie1a}), is taken with respect to the jump amplitude
${\bf J}_\theta(t)$.

The dimension of the (complex-valued) ODEs for ${\bf{\bar B}}({\bf u};\tau,\theta)$ corresponds to the dimension of the state vector ${\bf{X}}(t)$. Then, for stochastic parameter ${\bf \vartheta}$, the ChF is given by:
\begin{eqnarray*}
\phi_{{\bf X}}({\bf u};t,T):=\E_t[\e^{-\int_t^T r(s)\ds+i{\bf u}^\T{\bf X}(T)}]=\E_t\left[\E_t\big[\e^{-\int_t^T r(s)\ds+i{\bf u}^\T{\bf X}_\theta(T)}\big|{\bf \Theta}={\theta}\big]\right].
\end{eqnarray*}
The inner expectation can be recognized as the conditional ChF in~(\ref{eqn:ChFCond}); thus, by definition of the ChF and integration over all the parameter space, we find,
\begin{eqnarray}
\label{eqn:ChFIntegral}
\phi_{{\bf X}}({\bf u};t,T)=\E_t\left[\phi_{{\bf X}|{\Theta}}({\bf u};t,T)\right]
=\int_{\R^n} \phi_{{\bf X}|{\Theta}=\theta}({\bf u};t,T)f_{\Theta}(\theta)\d{\theta}.
\end{eqnarray}
We aim to provide numerically efficient methods for computation of ChF in~(\ref{eqn:ChFIntegral}). 

\begin{rem}[Simplified notation]
The representation above is rather generic in terms of multidimensional ChF and possibly a multidimensional set of stochastic parameters. However, such a generic case is rather exceptional, as, in a typical pricing application, a payoff function will only depend on a single, marginal distribution (even so, possibly, driven by the multidimensional system of the SDEs); therefore, for the sake of simplicity, from now on, we focus on the derivations of the Chf for ${\bf u}=[u,0,\dots]$, and we denote it by $u$. Moreover, although the presented framework allows for simultaneous randomization of multiple parameters, we will consider a single stochastic parameter $\vartheta$ with $\theta$ indicating its realization. In the following sections, we consider a finite number of realizations of $\vartheta$; therefore, we will denote them as $\theta_1,\dots,\theta_N$, for some $N\in\mathbb{N}$. These ``specific'' realizations we will interchangeably call either ``collocation''~\citep{scmc2019} or ``quadrature'' points. Finally, by $\vartheta(\hat a,\hat b, \cdot)$ we denote that $\theta$ is a random variable driven by parameters, $\hat a$, $\hat b$.
\end{rem}

To determine the ChF of an affine model with a {\it randomized parameter}, one needs to integrate the parameter's probability density function over the whole domain~(\ref{eqn:ChFIntegral}). This can be avoided, i.e., the complicated integrand can be factored into a set of pairs $\{\{\omega_1,\theta_1\},\dots,\{\omega_N,\theta_N\}\}$, $N\in\mathbb{N}$, with a nonnegative ``weights'' function, $\omega_i\geq0$, such that $\sum_{n=1}^N\omega_n=1$ and specific, collocation, points $\theta_n$. Once the number of evaluations, $N$, is low, we can significantly reduce the computational cost. The key element here, however, is that the pairs, $\{\omega_n,\theta_n\}_{n=1}^N$, cannot be chosen arbitrarily but need to be computed based on the parameter's distribution, $\vartheta$. We consider a random parameter $\vartheta$ with its PDF, $f_\vartheta(\cdot)$, such that for a fixed number  $N\in\mathbb{N}$, moments are finite, i.e., $\E[\vartheta^{2N}]<\infty$. Then we are able to determine a mapping, often called {\it quadrature rule}, function $\zeta(\vartheta):\R\rightarrow \{\omega_n,\theta_n\}_{n=1}^N$. We follow the approach presented by G.H. Golub and J.H. Welsch in~\citep{GoWe} where $\omega_n$ are the quadrature weights determined based on the moments of the random parameter, $\vartheta$. As explained in~\citep{GoWe}, one can calculate pairs
$\{\omega_n,\theta_n\}_{n=1}^N$ based on a three-term recurrence relation
which is well-known for orthogonal polynomials generated by $f_\vartheta(x)$, or equivalently, by the moments of $\vartheta$. Although the technique to compute these pairs of points is well-founded, for the readers' convenience, the theoretical aspects and the algorithmic details~\footnote{The accompanying Python and MATLAB codes are available at~\url{https://github.com/LechGrzelak/Randomization}} are included in~\ref{app:collocationTheory}. Throughout the article, the method to compute these points will be denoted as $\zeta(\vartheta):\R\rightarrow \{\omega_n,\theta_n\}_{n=1}^N$. 

The RAnD method for determining the ChF with a randomized model parameter is presented in the theorem below.  

\begin{theorem}[Characteristic Function for Randomized Affine Jump Diffusion Processes]
\label{prop:RAnDChF}
Consider a random variable $\vartheta$ defined on some finite domain $D_\vartheta:=[a,b]$, with its PDF, $f_\vartheta(x)$, CDF, $F_\vartheta(x)$ and a realization $\theta$, $\vartheta(\omega)=\theta$ such that for some $N\in\mathbb{N}$ the moments are finite, $\E[\vartheta^{2N}]<\infty$. Let ${\bf X}(t)$ represent an affine state vector with some constant parameter $\theta$. Assuming that the corresponding ChF, $\phi_{{\bf X}|\vartheta=\theta}(\cdot)$, is well defined and $2N$ times differentiable w.r.t $\theta$, the unconditional ChF for the randomized ${\bf X}$,  exists and is given by:
\begin{eqnarray}
\label{eqn:randChF}
\phi_{{\bf X}}({ u};t,T)=\sum_{n=1}^{N}\omega_n\phi_{{\bf X}|\vartheta=\theta_n}({ u};t,T)+\epsilon_N=\sum_{n=1}^{N}\omega_n\e^{\bar A({ u};\tau,\theta_n)+{\bf{\bar B^\T}}({ u};\tau,\theta_n){\bf{X}}(t)}+\epsilon_N,
\end{eqnarray}
with $\bar{A}(u;\tau,\theta_n)$, $\bar{{\bf B}}^\T(u;\tau,\theta_n)$ are defined in~(\ref{duffie1a}) and where \begin{equation}
\label{eqn:RAnDError}
\epsilon_N=\frac{1}{(2N)!}\frac{\partial^{2N}}{\partial\xi^{2N}}\phi_{{\bf X}|\vartheta=\xi}({ u};t,T),
\end{equation}
for $a<\xi<b$ and where the pairs $\{\omega_n,\theta_n\}_{n=1}^N$ are the Gauss-quadrature weights and the nodes based on the parameter distribution, $f_{\vartheta}(\cdot),$ determined by $\zeta(\vartheta):\R\rightarrow \{\omega_n,\theta_n\}_{n=1}^N$ defined in~\ref{app:collocationTheory}.
\begin{proof}
Starting with the definition of the ChF and conditioning on a parameter $\vartheta,$ we find:
\begin{eqnarray*}
\phi_{{\bf X}}({\bf u};t,T):=\E_t[\e^{-\int_t^T r(s)\ds+i{\bf u}^\T{\bf X}(T)}]=\E_t\left[\E_t\big[\e^{-\int_t^T r(s)\ds+i{\bf u}^\T{\bf X}_\theta(T)}\big|{\vartheta}={\theta}\big]\right].
\end{eqnarray*}
Then, by definition of the ChF,
\begin{eqnarray*}
\phi_{{\bf X}}({\bf u};t,T)&=&\E_t\left[\phi_{{\bf X}|\vartheta=\theta}({\bf u};t,T)\right]
=\int_\R \phi_{{\bf X}|\vartheta=\theta}({\bf u};t,T)f_\vartheta(x)\dx\\
&=&\sum_{n=1}^{N}\omega_n\phi_{{\bf X}|\vartheta=\theta_n}({\bf u};t,T)+\epsilon_N.
\end{eqnarray*}
First, we prove that a convex linear combination of characteristic functions, $\sum_{n=1}^{N}\omega_n\phi_{{\bf X}|\vartheta=\theta_n}({\bf u};t,T)$, and $\sum_{n=1}^N\omega_n=1$ with $\omega_n>0$ of a finite or a countable number of characteristic function is the characteristic function.

Since $\omega_n$ can be associated with probabilities, we can define a discrete random variable $\P[\vartheta=\theta_n]=\omega_n$, so we have:
\begin{eqnarray*}
\sum_{n=1}^{N}\omega_n\phi_{{\bf X}|\vartheta=\theta_n}({\bf u};t,T)=\sum_{n=1}^{N}\E[1_{\vartheta=\theta_n}]\E_t\big[\e^{i{\bf u}^\T{\bf X}_{\theta_n}(T)}\big]=\sum_{n=1}^{N}\E\big[1_{\vartheta=\theta_n}\e^{i{\bf u}^\T{\bf X}_{\theta_n}(T)}\big],
\end{eqnarray*}
since $\vartheta$ is independent of ${\bf X}(t)$. For Fubini's theorem to hold, the boundness condition needs to be satisfied, 
\begin{eqnarray*}
\sum_{n=1}^{N}\E\big[\big|1_{\vartheta=\theta_n}\e^{i{\bf u}^\T{\bf X}_{\theta_n}(T)}\big|\big]=\sum_{n=1}^{N}\E\big[1_{\vartheta=\theta_n}\big|\e^{i{\bf u}^\T{\bf X}_{\theta_n}(T)}\big|\big]\leq \sum_{n=1}^{N}\P\big[\vartheta=\theta_n\big]=1<\infty,
\end{eqnarray*}
therefore:
\begin{eqnarray*}
\sum_{n=1}^{N}\omega_n\phi_{{\bf X}|\vartheta=\theta_n}({\bf u};t,T)
=\E\big[\sum_{n=1}^{N}1_{\vartheta=\theta_n}\e^{i{\bf u}^\T{\bf X}_{\theta_n}(T)}\big]
=\E\big[\exp\big({i{\bf u}^\T\sum_{n=1}^{N}1_{\vartheta=\theta_n}{\bf X}_{\theta_n}(T)}\big)=:\phi_{Y}({\bf u}),
\end{eqnarray*}
with $\phi_{Y}(\cdot)$ being a ChF of $\sum_{n=1}^{N}1_{\vartheta=\theta_n}{\bf X}_{\theta_n}(T)$.

Using the fact that the collocation points $\theta_n$ for $n=1,\dots,N$ are optimal and correspond to the zeros of orthogonal polynomials, we relate the collocation method to Gaussian quadrature where the integral of a complex-valued function, $\phi_{{\bf X}|\vartheta=\theta}({\bf u};t,T)$,
can be approximated by a polynomial: \[\int_\R \phi_{{\bf X}|\vartheta=x}({\bf u};t,T)f_\vartheta(x)\dx-\sum_{n=1}^N \phi_{{\bf X}|\vartheta=\theta_n}({\bf u};t,T)
\omega_i=\frac{1}{(2N)!}\frac{\partial^{2N}}{\partial\xi^{2N}}\phi_{{\bf X}|p=\xi}({\bf u};t,T)=:\epsilon_N,\] 
with $f_\vartheta(x)$ the weight function, and $\omega_n$ the quadrature weights.  The error is given by $\epsilon_N$, for $a<\xi<b$, and the pair $\{\omega_n,\theta_n\}_{n=1}^N$ are the Gauss-quadrature weights and the nodes based on the parameter distribution
%\[\epsilon_N=\frac{1}{(2N)!}\frac{\partial^{2N}}{\partial\xi^{2N}}\phi_{{\bf X}|p=\xi}({\bf u};t,T),\] 
as proven in~\citep{bulirsch2002introduction} (p.180, Theorem 3.6.24).
\end{proof}
\end{theorem}

Theorem~\ref{prop:RAnDChF} illustrates that the ChF of the randomized AD model is a weighted sum of a set of conditional ChFs evaluated at certain realizations, $\theta_n$, of the underlying stochastic parameter $\vartheta$. The theorem also shows, in Equation~(\ref{eqn:RAnDError}), the exponential decay of the error in terms of $N$ associated with the quadrature approximation- suggesting that only a few terms will be needed to reach high-precision.

The key feature of the RAnD method is its ability to facilitate fast evaluation of the randomized ChF~(\ref{eqn:randChF}).  The construction of the ChF is based on the randomized variable, $\vartheta$, via its $2N$ moments, i.e., the availability of the moments allows us to determine $\zeta(\vartheta)$ which provides the quadrature points $\{\omega_n,\theta_n\}_{n=1}^N$.  It is, therefore, essential that the randomizing variables enable the closed-form moment calculation.  Moreover, when specifying a randomizing variable, its support must correspond to the physical range of the randomized parameter; for example,-- some of the model parameters may be only valid for specific ranges, such as correlation or volatility coefficients.  In Table~\ref{Tab:Moments}, a few possible distributions with their corresponding moments are provided.  When analytical moments are available, the computation of the corresponding points only requires the computation of a Cholesky decomposition and certain eigenvalues (see~\ref{app:collocationTheory}); it is therefore computationally cheap.
\begin{table}[htb!]
\centering\footnotesize
\caption{\footnotesize Selected distributions for the stochastic parameters. For the normal random variable with some mean, $\mu$, and variance, $\sigma^2$, it is sufficient to consider standard normal distribution, $\mathcal{N}(0,1)$, and properly scale the $\theta_n$ points, obtained from Algorithm~\ref{alg:collPoints}.}
\begin{tabular}{c|c|c}
name&raw moment&domain  \\\hline\hline
$\vartheta\sim\mathcal{U}([\hat a,\hat b])$&$\E[\vartheta^n]=\frac{\hat b^{n+1}-\hat a^{n+1}}{(n+1)(\hat b-\hat a)}$&$[\hat a,\hat b]$ \\
$\vartheta\sim \exp(\hat a)$&$\E[\vartheta^n]=\frac{n!}{\hat a^n}$& $\R^+$\\
$\vartheta\sim\mathcal{N}(0,1)$&$\E[\vartheta^n]=(n-1)!!$\;\text{if}\;\;$n$\;\text{even}; $0$ otherwise&$\R$ \\
$\vartheta\sim\Gamma(\hat a,\hat b)$&$\E[\vartheta^n]=\hat b^n\Gamma(n+\hat a)/\Gamma(\hat b)$&$\R^+$\\
$\vartheta\sim\chi^2(\hat a,\hat b)$&$\E[\vartheta^n]= 2^{n-1}(n-1)!(\hat a+n\hat b)+\sum_{j=1}^{n-1} \frac{(n-1)!2^{j-1}}{(n-j)!}(\hat a+j\hat b )\E[\vartheta^{n-j}]$&$\R^+\cup\{0\}$\\
\end{tabular}
\label{Tab:Moments}
\end{table}

%%%%%%%%%%%%%%%%%%%%%%%%%%%%%%%%%%%%%%%%%%%%%%%%%%%%%%%%%%%%%%%%%
\subsection{Pricing with the RAnD method}
%%%%%%%%%%%%%%%%%%%%%%%%%%%%%%%%%%%%%%%%%%%%%%%%%%%%%%%%%%%%%%%%%
We will focus on pricing two different plain vanilla options, namely European options on an asset, $S(t)$, and options on the associated volatility, VIX. The pricing will rely on a Fourier inversion method, namely the COS method~\citep{OosterleeGrzelakBook}, which, by a change of the pricing coefficients, $H_k$ and the corresponding ChF allows for pricing of both contracts. The generic pricing equation is given by:
\begin{equation}\label{v3}
V(t_0)=  \e^{-r(T-t_0)}\sideset{}{'}\sum_{k=0}^{N_c-1}
\Re\left[ \phi_{\bf X}\left(\frac{k\pi}{b-a};t_0,T\right)\exp\left(-i
k \pi \frac{a}{b-a}\right) \right] \cdot H_k+\epsilon_{c_1},
\end{equation}
where $N_c$ represents the number of expansion terms, $\Re(\cdot)$ is the real part and where $H_k$ for $k\geq0$ are known in closed-form coefficients corresponding to the payoff function. We will use the $H_k$ coefficients derived for European-style call/put options (see~\citep{OosterleeGrzelakBook} for details) and for options on VIX (presented later, in Section~\ref{sec:Bates}, Lemma~\ref{lem:HkVIX}). Parameters, $a$ and $b$ are the {\it tuning} parameters used to determine the integration range; while the error $\epsilon_{c_1}$, is exponentially decaying in $N_c$.

\begin{rem}[Inversion of the randomized ChF]
\label{rem:RandDensity}
The application of Fourier inversion to the randomized ChF in~(\ref{eqn:randChF}) yields,
\begin{eqnarray}
f_{{\bf X}}(x)=\frac{1}{2\pi}\int_{\R}\e^{-iux}\sum_{n=1}^{N}\omega_n\phi_{{\bf X}|\vartheta=\theta}({ u};t,T)\d u=\sum_{n=1}^{N}\omega_nf_{{\bf X}|\vartheta=\theta}(x).
\end{eqnarray}
Since $\omega_1+\dots+\omega_N=1$, $\omega_n\geq0$, for $n=1,\dots,N$, which implies that the density of the affine, randomized, system of SDEs, ${\bf X}(t)$ can be expressed as a, possibly multi-modal, mixture distribution. Although the exact representation for the density $f_{\bf X}(\cdot)$ may be challenging to derive analytically, one can use Fourier inversion to recover the corresponding density and utilize it for swift option pricing. 
\end{rem}

A natural extension of the RAnD method is to consider two parameters that are stochastic and follow a bivariate distribution. Such an extension would require that the conditional moments are known explicitly- which is somewhat troublesome. To this day, only a few distributions allow for exact moments. However, if we stay, for example, within the Gaussian world, such an extension to a 2D case is trivial. Corallary~\ref{cor:bivariate} gives us the general details.
\begin{cor}[Random parameters with bivariate distribution]
\label{cor:bivariate}
Under a bivariate distribution $\Theta=[\vartheta_1,\vartheta_2]$ with $\zeta(\vartheta_1)=\{\omega_{1,n},\theta_{1,n}\}_{n=1}^N$ and conditioned on $\zeta(\vartheta_2|\vartheta_1)=\{\omega_{2,n},\theta_{2,n}\}_{n=1}^N$, the randomized ChF is given by:
\begin{eqnarray}
\phi_{{\bf X}}({ u};t,T)=\sum_{n_1=1}^{N}\omega_{n_1}\sum_{n_2=1}^{N}\omega_{n_2}\phi_{{\bf X}|\vartheta_1=\theta_{n_1},\vartheta_2=\theta_{n_2}}({ u};t,T)+\tilde\epsilon_N,
\end{eqnarray}
where $N$ indicates the number of expansion terms, $\vartheta_2|\vartheta_1$ indicates a conditional random variable, and the remaining specification follows Theorem~\ref{prop:RAnDChF}.
\end{cor}

%%%%%%%%%%%%%%%%%%%%%%%%%%%%%%%%%%%%%%%%%%%%%%%%%%%%%%%%%%%%
\subsection{Illustrative case: the RAnD Black-Scholes model}
\label{sec:RAnD_BS}
%%%%%%%%%%%%%%%%%%%%%%%%%%%%%%%%%%%%%%%%%%%%%%%%%%%%%%%%%%%%
As the first example for the RAnD method, let us consider the randomized Black-Scholes (RAnD BS) model. Given the probability space $(\Omega,\F(t),\Q)$, we define a stock $ S(t):=\{S(t,\omega):t\in T\}$ and the randomizing volatility $\sigma:=\{\sigma(\omega):\Omega\rightarrow \R^+\}$, with some constant interest rate, $r$. We  consider three different, continuous, randomizers for $\sigma$, i.e., we assume the volatility parameter, $\sigma$, to either follow a uniform distribution, $\sigma\sim\mathcal{U}([\cdot,\cdot])$, gamma distribution, $\sigma\sim\Gamma(\cdot,\cdot)$, or the non-central chi-squared distribution, $\sigma\sim\chi^2\left(\cdot,\cdot\right)$; thus, also mimicking a similar structure as in the Heston model where the CIR dynamics drive the variance process~(see Equations~(\ref{eqn:varianceHeston}) and~(\ref{eqn:non-central}) for the exact relation). 

Formally, the RAnD BS model is given by the following dynamics:
\begin{eqnarray*}
\d S(t)&=&r S(t)\dt + \sigma S(t)\dW(t),\;\;S(t_0)>0,\\
\sigma&\sim&\mathcal{U}([\hat a,\hat b]),\;\;\;\;{\text{or}}\;\;\;\;\sigma\sim\Gamma(\hat a,\hat b),\;\;\;\;\text{or}\;\;\;\;\sigma\sim\hat a\chi^2(\hat b,\hat c),
\end{eqnarray*}
for $\hat{a},\hat{b},\hat{c}\in \R^+.$ As it will be presented later, the randomization of the BS model is elegant as it allows for a generation of the European option prices in closed form (see~\cite{brigo2002lognormal} where discrete randomizers for the BS model were covered). However, since the randomized volatility is time-invariant, or {\it stationary}, such a model cannot compete with, e.g. the model of Heston, where the volatility has a term structure, and it is driven by a correlated stochastic process. Contrary to the Heston model, however, the RAnD BS model does not require Fourier transformation to price options. 

We have two techniques to compute option prices of the randomized Black-Scholes model. First, a straightforward approach is based on the closed-form solution of the ``unrandomized'' model. The second method is based on randomization at the ChF level. By application of the RAnD method and conditioning on the volatility realizations, the price of a European call option, for $t_0=0$, reads:
\begin{eqnarray*}
V(t_0)&=&\e^{-rT}\E_{t_0}\left[\max\left(S(T)-K,0\right)\right]\\
&=&\e^{-rT}\E_{t_0}\left[\E_{t_0}\left[\max\left(S(T)-K,0\right)\big|\sigma=\sigma_n\right]\right]=\e^{-rT}\E_{t_0}\left[V_c(t_0,T,K,S(t_0),r,\sigma_n)\right],
\end{eqnarray*}
with $\E_{t_0}[\cdot]$ being the expectation operator with filtration $\F(t_0)$, 
where $V_c(t_0,T,K,S(t_0),r,\sigma_n)$ represents a European call option price with strike $K$, expiry $T$ and the corresponding interest rate $r$, and it is defined in~(\ref{eqn:BS}) below. The price $V(t_0)$ is now given by:
\begin{eqnarray}
\nonumber
V(t_0)&=&\e^{-rT}\int_{\R^+}V_c(t_0,T,K,S(t_0),r,x) f_{\sigma}(x)\dx\\
\label{eqn:BS_sigmaN}
&=&\e^{-rT}\sum_{n=1}^N\omega_nV_c(t_0,T,K,S(t_0),r,\sigma_n)+\hat\epsilon_N,
\end{eqnarray}
where pairs $\{\omega_n,\sigma_n\}$ are the weights and the quadrature points, determined based on the distribution of $\sigma(\omega)$ and are computed using Algorithm~\ref{alg:collPoints} in~\ref{app:collocationTheory}, $\hat\epsilon_N$ is the quadrature error defined in~(\ref{eqn:randChF}).

At each realization $\sigma_n,$ the pricing is known in closed form and is given the Black-Scholes formula:
\begin{eqnarray}
\label{eqn:BS}
V_c(t_0,T,K,S(t_0),r,\sigma_n)=S(t)F_{\mathcal{N}(0,1)}(d_1)-KF_{\mathcal{N}(0,1)}(d_2)\e^{-r(T-t_0)},
\end{eqnarray}
with $d_1=\frac{1}{\sqrt{\sigma_n^2(T-t_0)}}\left[\log S(t_0)/K+\left(r+1/2\sigma_n^2\right)(T-t_0)\right],\;\;\;d_2=d_1-\sqrt{\sigma_n^2(T-t_0)}.$

\begin{rem}[Continuous randomizers for the BS model- previous attempts]
Early attempts to randomize the Black-Scholes model using continuous random variables are already known in the literature. In~\citep{jacquier2015black}, the authors have randomized the variance of the Black-Scholes model by a CEV-generated distribution. The primary aim of such randomization was to achieve the so-called {\it moderately explosive} distribution allowing for enhanced skew. The presented framework, however, although via the asymptotic expansion allowed for short-maturity proxies for the implied volatilities, is very much problem and randomizer specific. Moreover, due to the nature of the expansion, it performs well only for short-expiry derivatives, contrary to the RAnD method discussed in this article.
\end{rem}

%%%%%%%%%%%%%%%%%%%%%%%%%%%%%%%%%%%%%%%%%%%%%%%%%%%%%%%%%%%%%%%%%%%%%%%%%%%%%%%%%%%
\subsubsection{Randomization of the ChF under the Black-Scholes model}
%%%%%%%%%%%%%%%%%%%%%%%%%%%%%%%%%%%%%%%%%%%%%%%%%%%%%%%%%%%%%%%%%%%%%%%%%%%%%%%%%%%

The presented randomization technique is not limited to the Black-Scholes model, where a closed-form solution exists for each volatility realization, $\sigma_n$. In order to achieve more flexibility and be able to price challenging derivatives, the stochastic volatility models, like the Heston and Bates models, will be considered. A numerical routine, like Fourier inversion, needs to be used. Although such inversions are swift due to exponentially convergent methods, like the COS method~\citep{OosterleeGrzelakBook}, it is important to keep the number of inversions as low as possible. This requirement suggests that a preferred way to compute a randomized option price is via randomization at the ChF level, introduced in~(\ref{eqn:randChF}), which for the RAnD BS model is given by the following expression,
\begin{eqnarray}
\label{eqn:ChF_BS}
\phi_{{X}}({ u};t_0,T)=\sum_{n=1}^{N}\omega_n\exp\big({iuX(t_0)+(r-\frac12\sigma_n^2)iu(T-t_0)-\frac12\sigma_n^2u^2(T-t_0)}\big)+\epsilon_N,
\end{eqnarray}
with $X(t_0)=\log S(t_0)$, the error $\epsilon_N$ defined in~(\ref{eqn:randChF}). 

The pricing of European-style options can be performed using the expansion formula in~(\ref{v3}).

At this point, one can still employ the pricing equation~(\ref{v3}) to~(\ref{eqn:BS_sigmaN}); however, it would require $N$ evaluations of the pricing equation in~(\ref{v3}). Since the strength of the COS method is its ability to compute option prices for a range of strikes $K$, it is important to keep the number of vector-matrix multiplications in~(\ref{v3}) to a minimum; it is, therefore, beneficial to evaluate the pricing only once- thus the randomization should take place at the ChF level (as a ChF does not depend on strikes). An additional element that needs to be considered is the generated error, i.e., depending on the smoothness of the option value~(\ref{eqn:BS_sigmaN}) or the corresponding ChF~(\ref{eqn:ChF_BS}), error $\hat\epsilon_N$ may differ from the $\epsilon_N$ in~(\ref{eqn:BS_sigmaN}).

%%%%%%%%%%%%%%%%%%%%%%%%%%%%%%%%%%%%%%%%%%%%%%%%%%%%%%%%%%%%%%%%%%%%%%%%%%%%%%
\subsubsection{RAnD BS model: convergence and implied volatilities}
\begin{figure}[h!]
  \centering
      \includegraphics[width=0.45\textwidth]{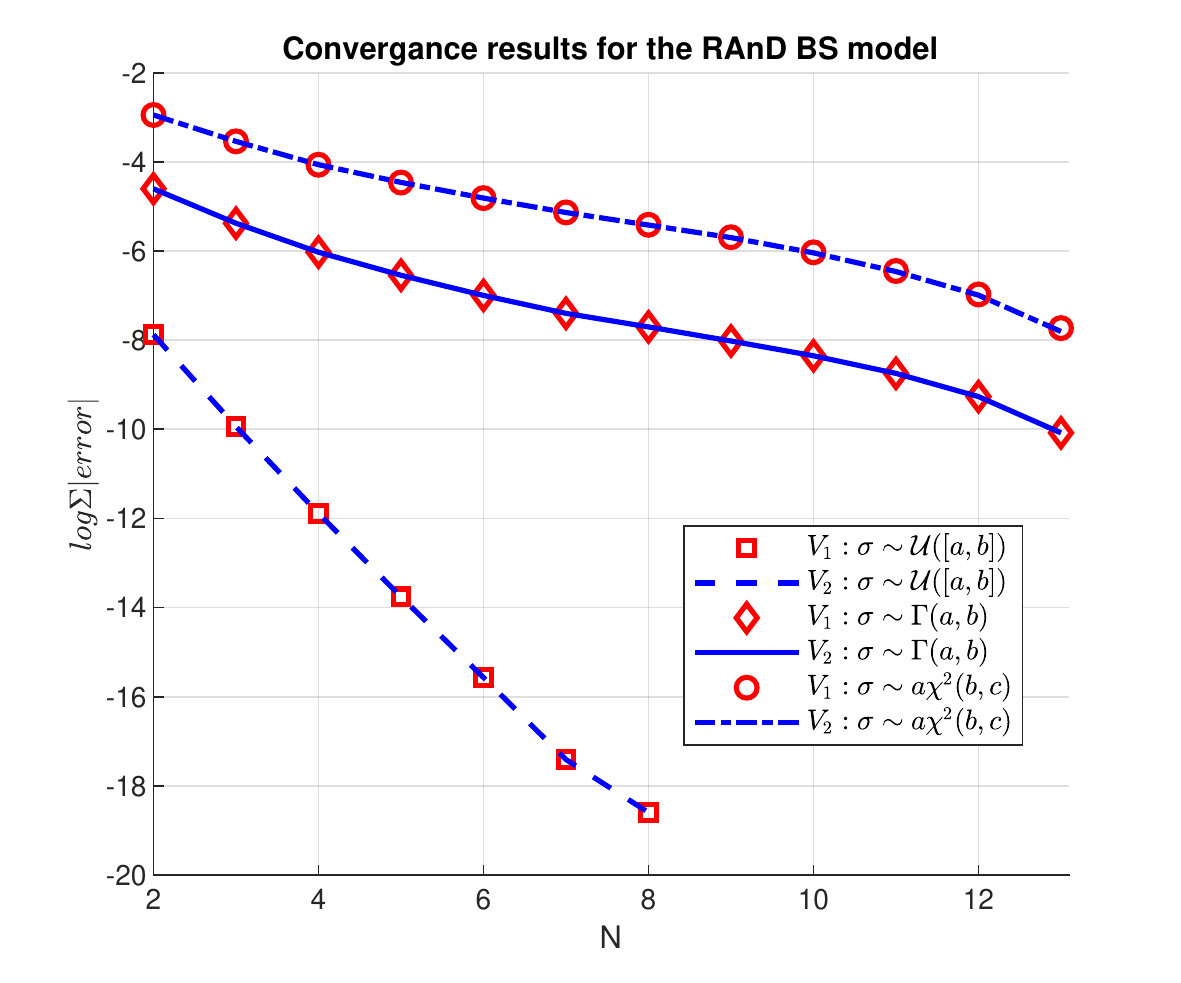}
      \includegraphics[width=0.45\textwidth]{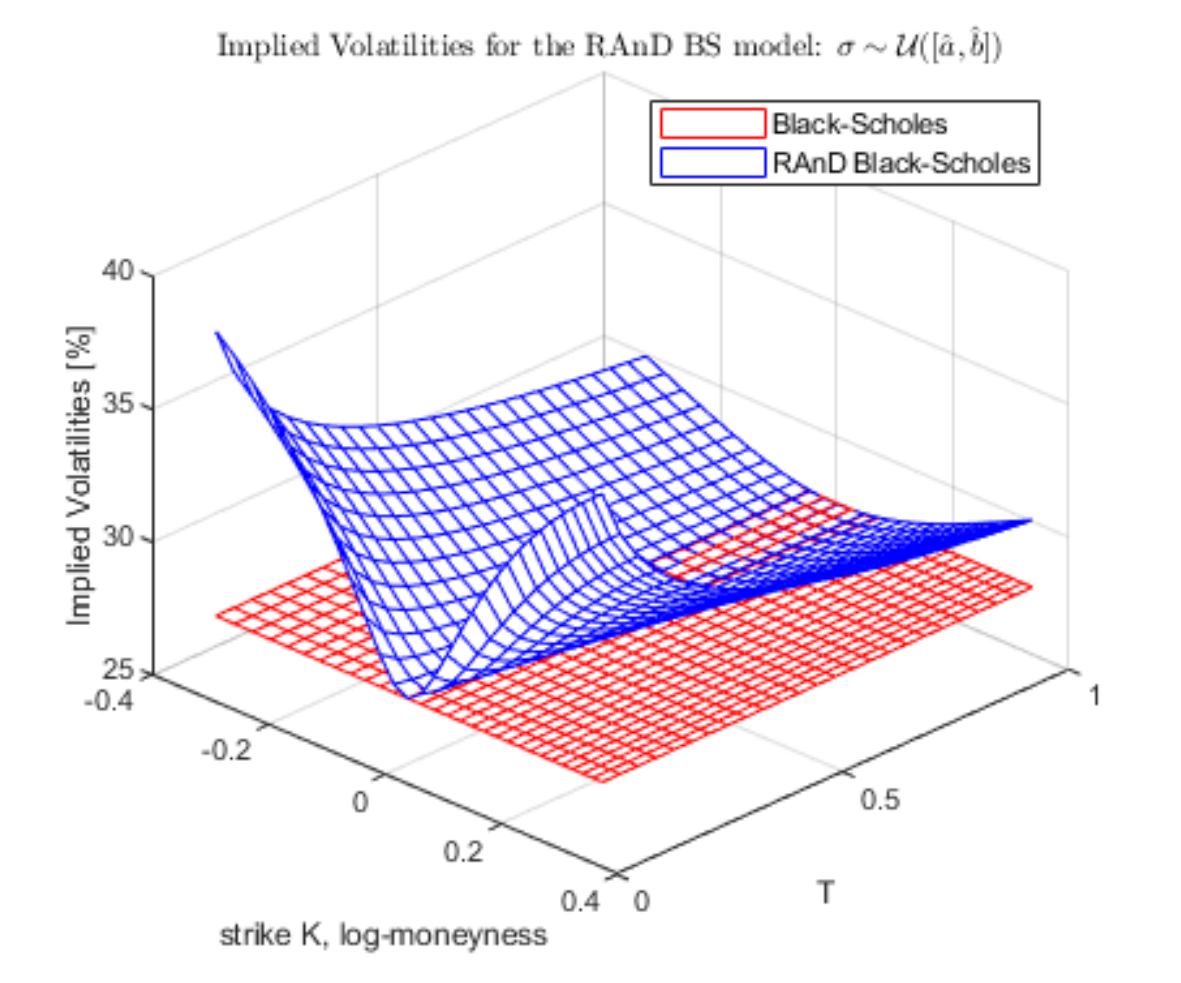}
          \caption{Left: The RAnD method convergence for the number of collocation points $N$. Red markers correspond to the results obtained based on analytical expression~(\ref{eqn:BS_sigmaN}), and blue lines correspond to Fourier expansion~(\ref{eqn:ChF_BS}). The ``error'' is the difference between option prices against the reference, fetched with 5 million paths. The summation is over the entire range of option strikes. Right: Implied volatility surface for the RAnD BS model for ${\sigma\sim\mathcal{U}([\hat a,\hat b])}$.  }
      \label{fig:RandBS}
\end{figure}

%%%%%%%%%%%%%%%%%%%%%%%%%%%%%%%%%%%%%%%%%%%%%%%%%%%%%%%%%%%%%%%%%%%%%%%%%%%%%%
In order to assess the convergence of the RAnD method and the impact of the randomization on implied volatilities, we consider a numerical experiment, where two pricing approaches, presented in the previous section, will be compared. In particular, we are interested in the implied volatility behaviour for short maturity options ($T=0.1$), as this is often considered a challenging case. We start by specifying the parameters for the randomizing random variables (see Table~\ref{Tab:parametersExperiment}). 
\begin{table}[htb!]
\centering\footnotesize
\caption{\footnotesize Parameters used in the RAnD BS model simulation. For the non-central chi-square distributions, the parameters were chosen based on the Heston model (see Dynamics~(\ref{eqn:Bates})), with: $\gamma=1.9$, $\kappa=0.5$, $\bar{v}=0.3$, $v_0=0.3$, $T=0.1.$ }
\begin{tabular}{c|c|c|c}
&$\hat a$&$\hat b$&$\hat c$  \\\hline\hline
$\sigma\sim\mathcal{U}([\hat a,\hat b])$&0.1&0.45&- \\
$\sigma\sim\Gamma(\hat a,\hat b)$&2.55&0.1&-\\
$\sigma\sim\hat a\chi^2(\hat b,\hat c)$&0.088&0.1662&3.2417\\
\end{tabular}
\label{Tab:parametersExperiment}
\end{table}
With the distributions given, we are able to determine the corresponding pairs $\{\omega_n,\sigma_n\},$ for $n=1,\dots,N$. These points are based on the moments of the randomizing random variable (see Table~\ref{Tab:Moments}) and utilization of Algorithm~\ref{alg:collPoints}. %For replication purposes, these points are tabulated in Table~\ref{Tab:CollPoints_probabilitiesUniform} in~\ref{sec:collPoints_Randomizer_appendix}. 
Figure~\ref{fig:RandBS} (left) illustrates the method's convergence regarding the number of collocation points $N$. Surprisingly, for the randomized Black-Scholes model, it does not matter whether the randomization takes place at the price level or the corresponding ChF. However, the convergence rate is affected by the associated randomizer. We report the highest convergence speed for the uniform distribution, lower for gamma, and the slowest convergence rate for the non-central chi-square randomizer. The phenomenon can be explained by the {\it amount} of randomness introduced to the pricing equation.  

In Figure~\ref{fig:RandBS} (right), the impact of $\sigma\sim\mathcal{U}([\hat{a},\hat{b}]),$  on the implied volatilities is illustrated (the remaining cases are illustrated in~\ref{app:randomizersImpact} in Figure~\ref{fig:RandBS_2}). As expected, the RAnD BS model can generate an implied volatility smile. However, because of the independence between the stock's Brownian motion and the randomizer, the skew control is limited, irrespective of the randomizing distribution. It is important to note that by introducing stochastic but time-invariant volatility, we can generate an implied volatility surface that develops in time from a very peaked, {\it almost explosive}, to a flattening shape. The implied volatility surface pattern is very similar in all three cases. The RAnD BS model could be further extended, with, for example, a displacement parameter that would introduce skewness in the model, thus further extending the applicability of the Black-Scholes model's standard. 

In the final experiment for the RAnD BS model, we report the convergence rate in terms of the implied volatilities in Table~\ref{Tab:RAnd_ImpliedVols}. We see that the convergence is not affected by the maturity time $T$. We expect that, with stochastic volatility models, the convergence will depend on the time-to-expiry. The results are excellent; for $N=6$ points, the maximum absolute error is about $0.05-0.07\%$, which is a convincing result and implies that the method can be used for accurate model calibration and pricing. As expected, the reported errors depend on the randomizing variable (compare Table~\ref{Tab:RAnd_ImpliedVols} against Table~\ref{Tab:RAnd_ImpliedVolsUniform} in~\ref{sec:RAnDBS_uniform}). 

\begin{table}[htb!]
\centering\footnotesize
\caption{\footnotesize Maximum implied volatility, $\text{IV}(\cdot)$, error computed against Monte Carlo results with 10 million samples, with $K=S_0\e^{0.1\sqrt{T}\delta}$, $\delta=[-3,-2,-1,-0.5,0,0.5,1,2,3]$.}
\begin{tabular}{c|c|c|c|c|c|c|c|c}
\multicolumn{8}{c}{$\text{error} =\max_i{\big|\text{IV}(K_i)-\text{IV}_{\text{ref}}(K_i)\big|}$}\\
$\sigma\sim\Gamma(a,b)$&$N=2$&$N=3$&$N=4$&$N=5$&$N=6$&$N=7$&$N=8$&$N=9$\\\hline\hline
$T=1d$&	$0.56\; \%$&$0.26\; \%$&$0.15\; \%$&$0.10\; \%$&$0.07\; \%$&$0.06\; \%$& $0.05\; \%$& $0.04\; \%$\\
$T=1w$& $0.56\; \%$&$0.26\; \%$&$0.15\; \%$&$0.10\; \%$&$0.07\; \%$&$0.06\; \%$& $0.04\; \%$& $0.04\; \%$\\
$T=2w$& $0.55\; \%$&$0.25\; \%$&$0.14\; \%$&$0.09\; \%$&$0.06\; \%$&$0.04\; \%$& $0.03\; \%$& $0.03\; \%$\\
$T=1m$& $0.54\; \%$&$0.24\; \%$&$0.13\; \%$&$0.08\; \%$&$0.06\; \%$&$0.04\; \%$& $0.03\; \%$& $0.03\; \%$\\
$T=3m$& $0.54\; \%$&$0.24\; \%$&$0.13\; \%$&$0.08\; \%$&$0.05\; \%$&$0.04\; \%$& $0.03\; \%$& $0.02\; \%$\\
$T=6m$& $0.55\; \%$&$0.25\; \%$&$0.14\; \%$&$0.09\; \%$&$0.07\; \%$&$0.05\; \%$& $0.04\; \%$& $0.03\; \%$\\
$T=12m$& $0.56\; \%$&$0.26\; \%$&$0.15\; \%$&$0.10\; \%$&$0.07\; \%$&$0.05\; \%$& $0.04\; \%$& $0.04\; \%$\\
\end{tabular}
\label{Tab:RAnd_ImpliedVols}
\end{table}
%\multicolumn{4}{c}{Smolyak grid}

Further analysis regarding randomization and implied volatility shapes will be discussed later in the context of the Heston and the Bates models.
%%%%%%%%%%%%%%%%%%%%%%%%%%%%%%%%%%%%%%%%%%%%%%%%%%%%%%%%%%%%
\section{The RAnD framework and pricing of plain vanilla and VIX options}
\label{sec:Bates}
%%%%%%%%%%%%%%%%%%%%%%%%%%%%%%%%%%%%%%%%%%%%%%%%%%%%%%%%%%%
This section focuses on applying the RAnD method to stochastic volatility and jump models. In particular, we will consider the Bates~\cite{Bates:1996} model, which forms an elegant extension of the model of Heston~\cite{Heston:1993}. From the pricing perspective, the characteristic function of both models is known analytically. Although some may argue that managing jumps in the pricing framework is of limited value, especially for hedging, it is well known that {\it pure} stochastic volatility, affine, models cannot generate sufficient skew for short maturity options. Adding jumps to calibrate short-maturity options is often a preferred choice. However, it is often insufficient. Our goal is to use the RAnD method and apply it to the challenging task of simultaneous pricing of plain-vanilla and VIX options. In particular, we illustrate the ability to fit both option markets simultaneously by randomizing some Bates model parameters. 

Randomization of individual Heston model parameters is known in the literature. Initial random volatility was discussed in~\cite{pages2020stationary,jacquier2019randomized,mechkov2015hot}, and in~\cite{fouque2018heston}, the perturbation-based approximations for stochastic {\it vol-of-vol} parameter were derived. 

We will focus on the Bates model: the corresponding ChF, pricing of plain-vanilla and VIX options and calibration, under the RAnD method, to market data.

The Bates model, under the $\Q$ measure, is described by the following system of SDEs:
\begin{eqnarray}
\label{eqn:Bates}
\begin{array}{l}
		\d S(t)/ S(t)=\left(r-\lambda\E[\e^J-1]\right)\dt+\sqrt{v(t)}\dW_x(t)+\left(\e^J-1\right)\d X_{\mathcal{P}}(t),\\[1.5ex]
\label{eqn:varianceHeston}\d
v(t)=\kappa\left(\bar{v}-v(t)\right)\dt+\gamma\sqrt{v(t)}\dW_v(t),
	\end{array}
\end{eqnarray}
with Poisson process $X_\mathcal{P}(t)$, intensity $\lambda$, and
normally distributed jump sizes,
$J\sim\mathcal{N}(\mu_j,\sigma^2_j)$, with $\E[\e^{J}]=\e^{\mu_J+\frac12\sigma_J^2}.$ 
$X_{\mathcal{P}}(t)$ is assumed to be independent of the Brownian motions and the jump sizes. There is a correlation $\rho$ between the governing Brownian motions, $\rho\dt =\dW_x(t)\dW_v(t)$.
Under this model the variance process follows the non-central chi-square distribution, $\chi^2(\delta,\bar \kappa(\cdot,\cdot))$ with $\delta$ degrees of freedom and non centrality parameter $\bar \kappa(t_0,t)$,
\begin{eqnarray}
\label{eqn:non-central}
v(t)|v(t_0)\sim \bar c(t_0,t) \chi^2(\delta,\bar \kappa(t_0,t)),
\end{eqnarray}
where
\begin{eqnarray}
\label{eqn:non-central_params}
	\bar c(t_0,t)=\frac{\gamma^2}{4\kappa}(1-\e^{-\kappa (t-t_0)}),\;\;\;
\delta=\frac{4\kappa\bar{v}}{\gamma^2},\;\;\;	\bar \kappa(t_{0},t)=\frac{4\kappa\e^{-\kappa (t-t_0)}v(t_0)}{\gamma^2(1-\e^{-\kappa (t-t_0)})}.
\end{eqnarray}
Under a log transformation, the dynamics of the stock, $X(t)=\log S(t),$ belongs to the class of affine jump-diffusion processes, and it reads:
\begin{eqnarray*}
	\begin{array}{l}
		\d X(t)=\left(r-\frac12 v(t)-\lambda\E[\e^J-1]\right)\dt+\sqrt{v(t)}\dW_x(t)+J\d X_{\mathcal{P}}(t),
	\end{array}
\end{eqnarray*}
so that, given Theorem~\ref{prop:RAnDChF}, we can derive the corresponding, randomized, ChF, which is given by the following expression:
\begin{eqnarray}
\label{eqn:rand_Bates}
\phi_{X}({ u};t,T)=\sum_{n=1}^{N}\omega_n\e^{iuX(t) +\bar{C}(u,\tau;\theta_n)v(t)+\bar{A}_B({u},\tau;\theta_n)}+\epsilon_N,
\end{eqnarray}
for $\tau=T-t$, with complex valued functions $\bar{A}_B(u;\tau,\theta_n)$ and $\bar{C}(u;\tau,\theta_m)$, given in~\ref{sec:AC}. The third argument $\theta_n$ represents a particular realization of model parameter, $\vartheta\in\{v_0,\lambda,\gamma,\rho,\mu_J,\sigma_J,\kappa,\bar{v}\}$, that we wish to randomize. 

\begin{figure}[h!]
  \centering
    \includegraphics[width=0.45\textwidth]{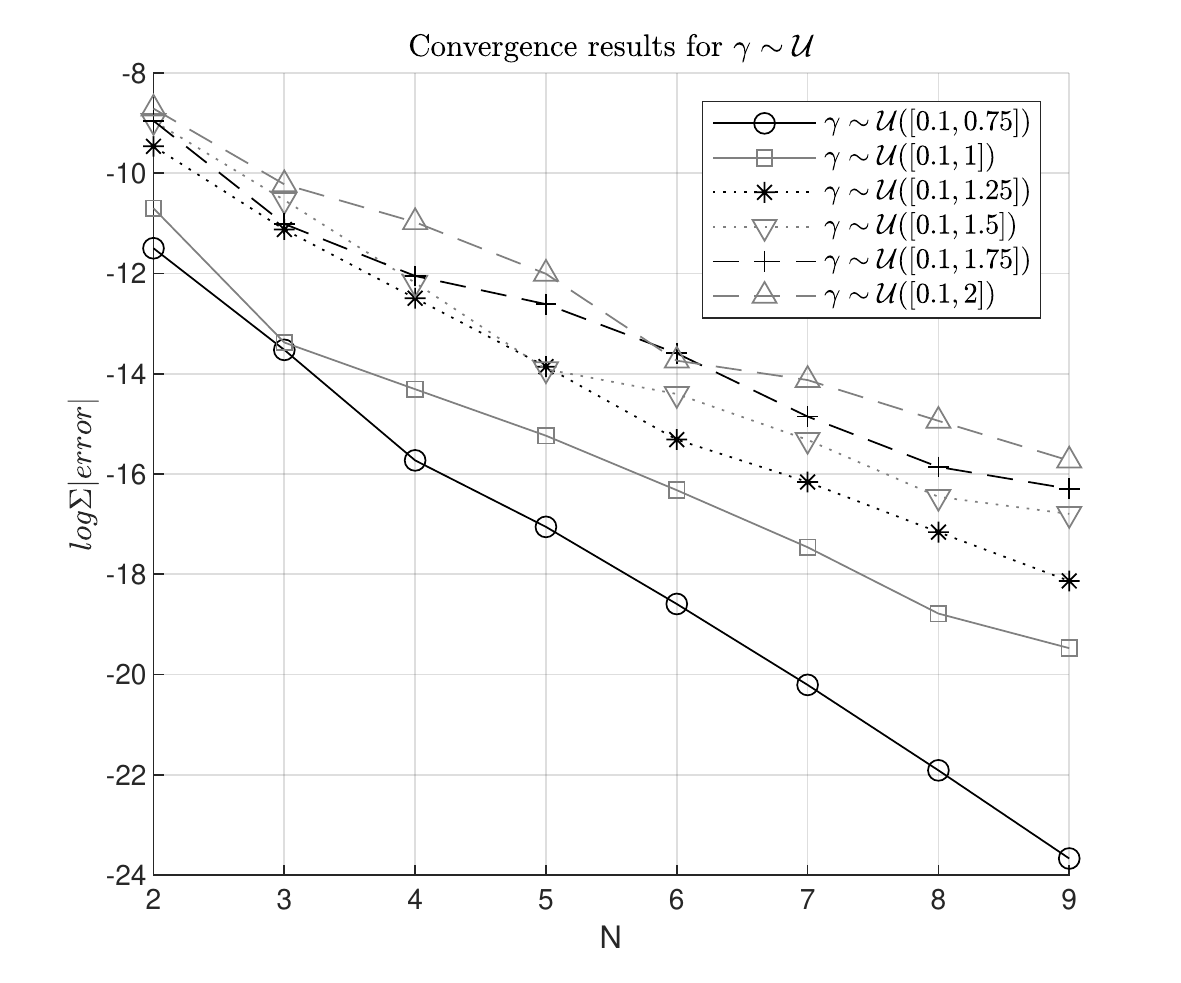}
    \includegraphics[width=0.45\textwidth]{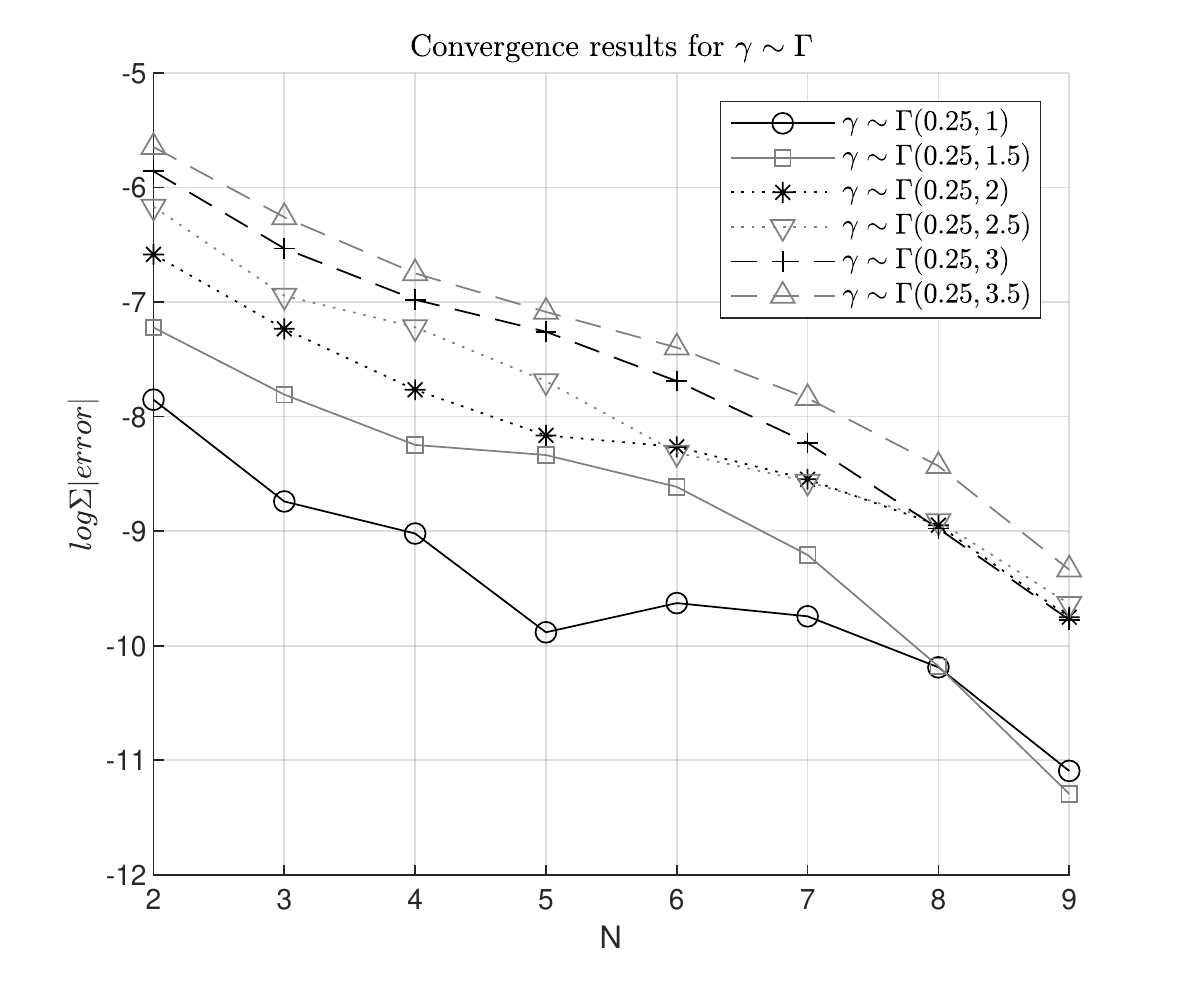}
      \caption{Convergence results for randomization of vol-of-vol parameter, $\gamma$ under the Bates model with: $T=31/365$, $r=0$, $\mu_J=-0.25$, $\sigma_J= 0.05$, $\lambda= 0.1$, $\kappa= 0.5$, $\gamma= 0.72$, $\bar{v}=0.1$, $\rho= -0.85$ and $v_0= 0.0625$.  Left: $\gamma\sim\mathcal{U}(\hat a,\hat b)$ for varying $\hat a$ and $\hat b$; Right: $\gamma\sim\Gamma(\hat a,\hat b)$ for varying $\hat a$ and $\hat b$.}
      \label{fig:convergence_gamma}
\end{figure}

Once the RAnD ChF is derived, it is worth checking the convergence concerning the number of summation terms, $N$, in~(\ref{eqn:rand_Bates}). As an example, in Figure~\ref{fig:convergence_gamma} the convergence for randomization of the vol-vol parameter, $\gamma$, with two distributions, uniform, $\mathcal{U}$, and gamma distribution, $\Gamma$, is depicted. The convergence rate depends on the distribution itself but also on its parameters. In both cases, the convergence rate is excellent; however, as expected, it is much faster for a uniform distribution than for the gamma distribution. In both cases, satisfactory results will be accomplished for $N=5$. Randomization of other model parameters has shown equivalent convergence patterns.

\begin{figure}[h!]
  \centering
    \includegraphics[width=0.45\textwidth]{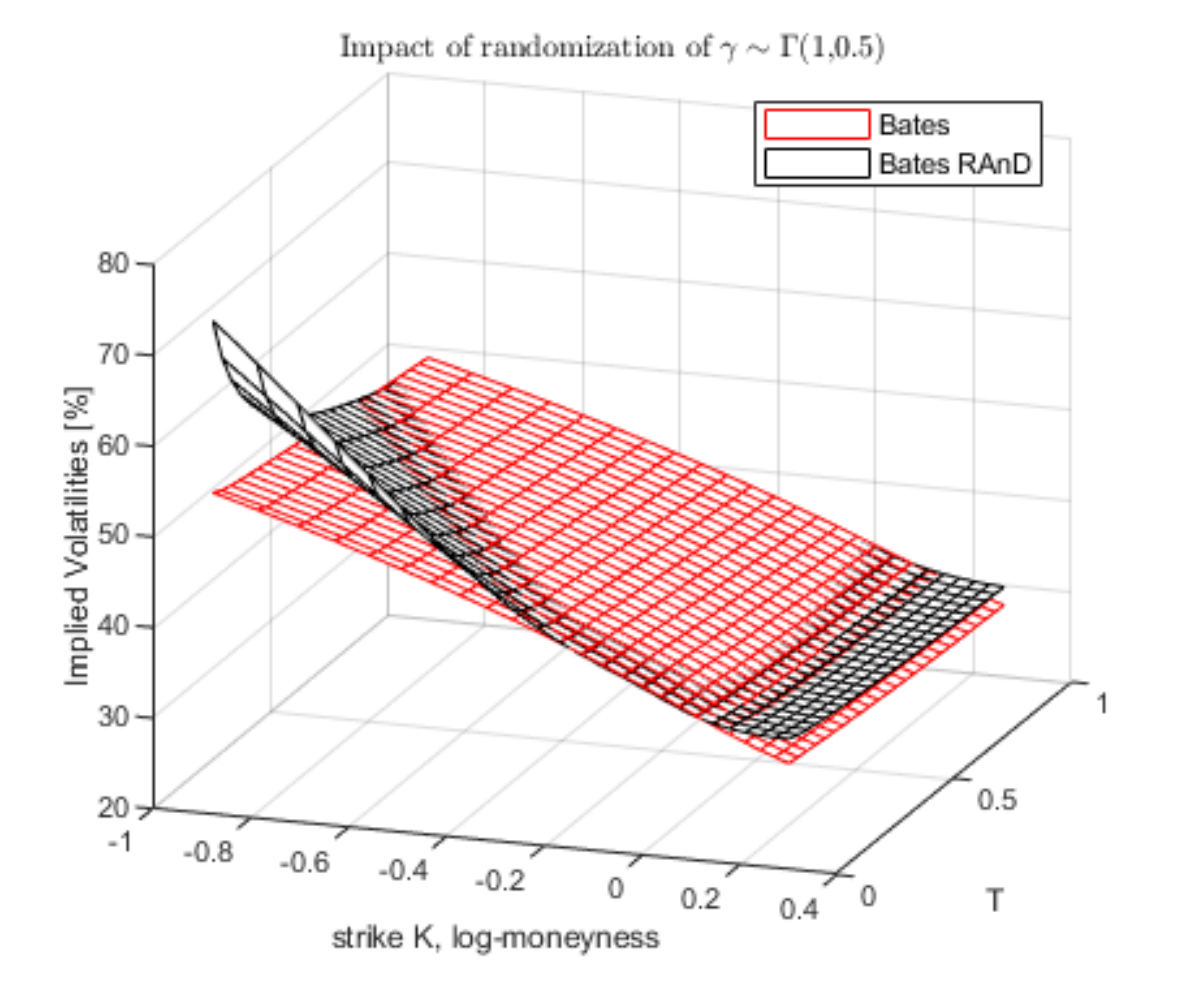}
    \includegraphics[width=0.45\textwidth]{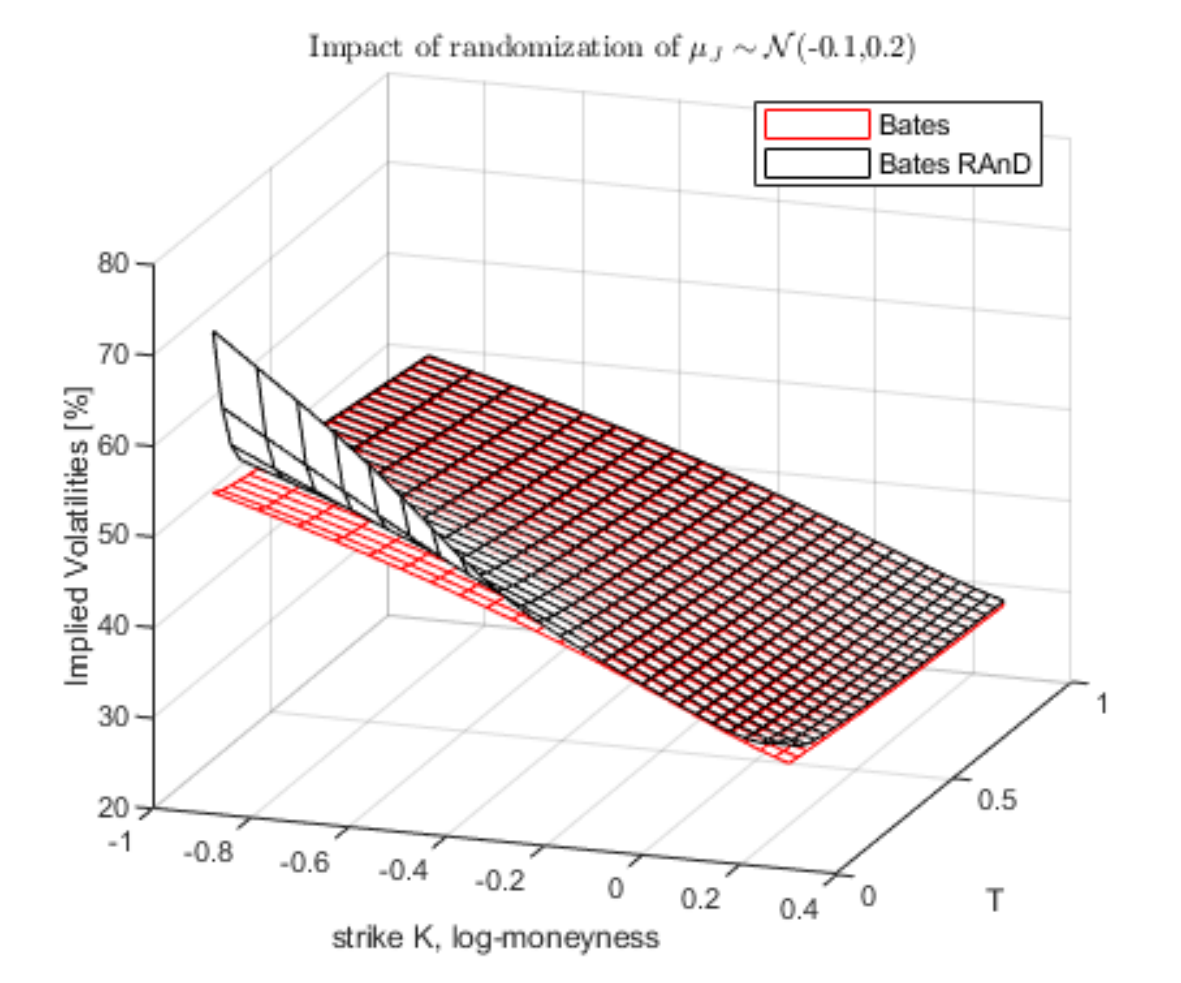}
      \caption{Implied volatility surface for RAnD Bates model. Left panel: randomized {\it vol-vol}, $\gamma~\sim\Gamma(1,0.5)$. Right panel: randomized {jump's mean}, $\mu_J~\sim\mathcal{N}(-0.1,0.2)$. Other model parameters are $r= 0$, $\mu_J = -0.1$, $\sigma_J= 0.06$, $\lambda = 0.08$, $\kappa     = 0.5$, $\gamma     = 0.5$, 
$\bar{v} = 0.13$, $\rho = -0.7$, $T         = 1/12$, and $ v_0 = 0.13$. }
      \label{fig:impact3D}
\end{figure}

The impact of the randomization for the Bates model can be analyzed with the ChF in~(\ref{eqn:rand_Bates}). In particular, the time evolution of the implied volatility surface for randomized vol-vol, $\gamma$, and the jump amplitude parameter, $\mu_J$, is presented in Figure~\ref{fig:impact3D}. The results are insightful, showing that the RAnD method can magnify the skewness of short-maturity options. Intriguingly, the evolution of the skew differs for both parameters, i.e., randomized $\gamma$ allows for slow skew decay towards the benchmark- the basic Bates model, while the randomization of $\mu_J$ generates much more rapid decay. Depending on the market data, both shapes may be desired.

\begin{figure}[h!]
  \centering
    \includegraphics[width=0.45\textwidth]{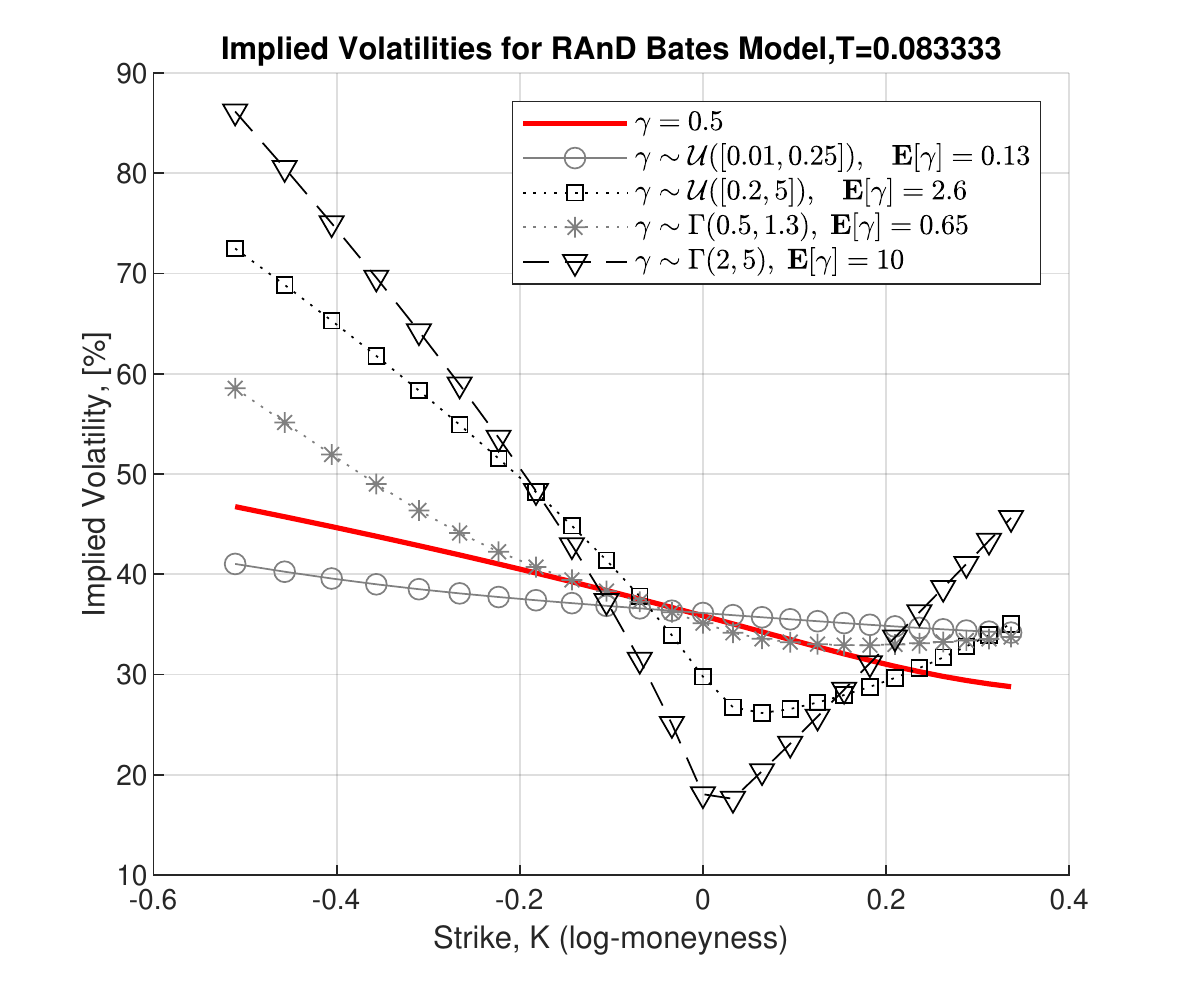}
    \includegraphics[width=0.45\textwidth]{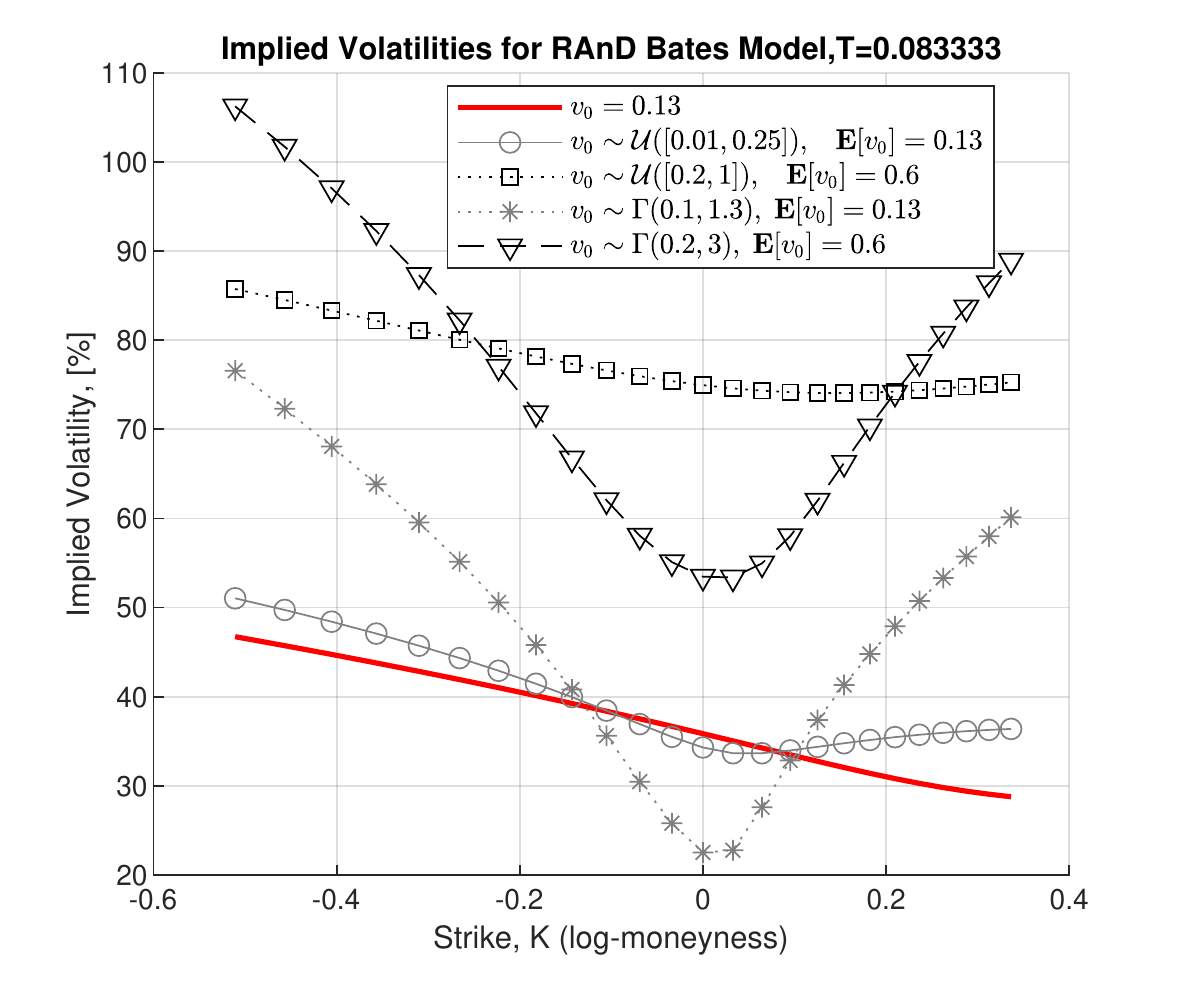}
      \caption{Impact of randomized parameters on implied volatilities. Left: randomized {\it vol-vol}, $\gamma$. Right: randomized {\it initial vol}, $v_0$.}
      \label{fig:impact1}
\end{figure}

\begin{figure}[h!]
  \centering
    \includegraphics[width=0.45\textwidth]{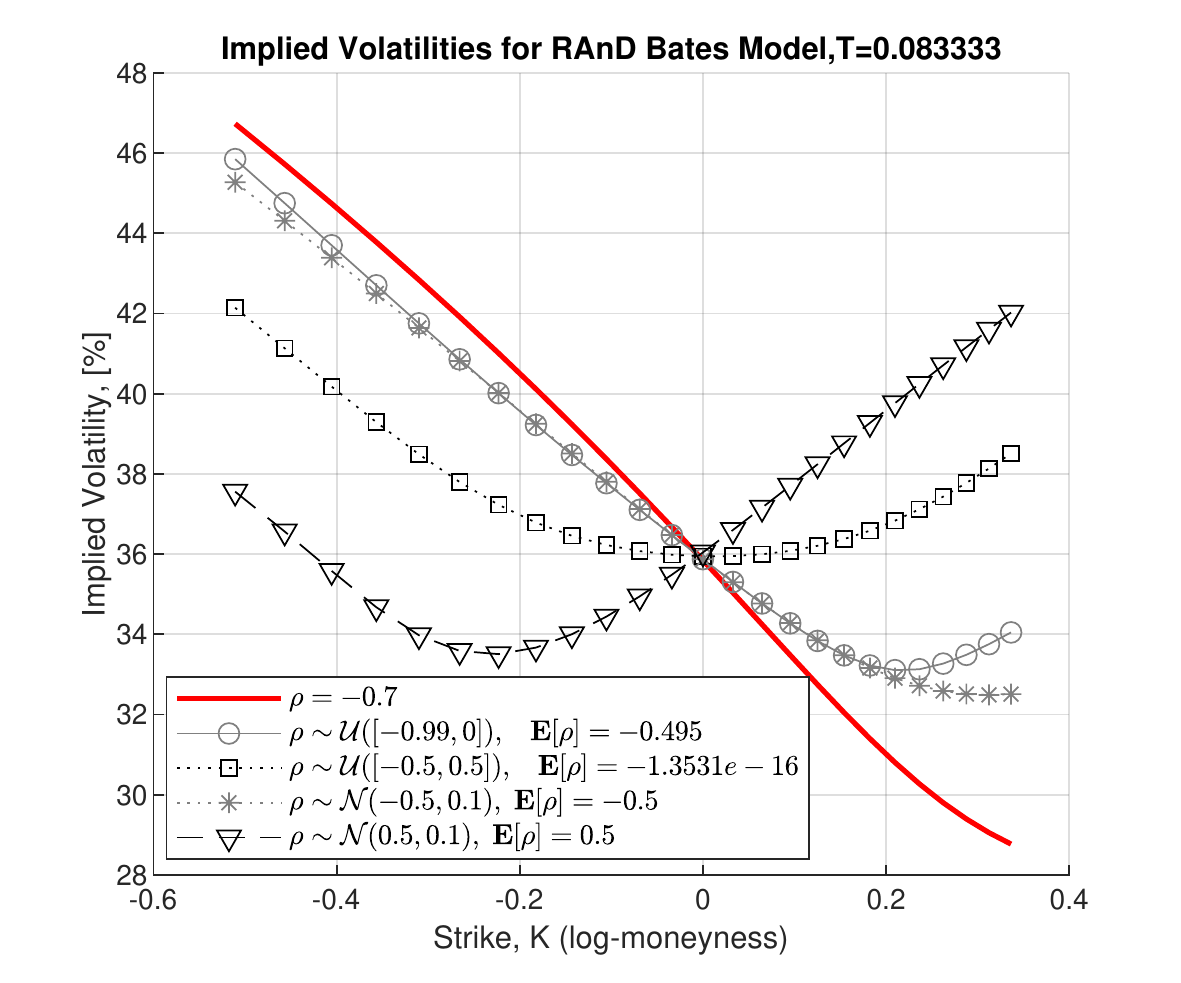}
    \includegraphics[width=0.45\textwidth]{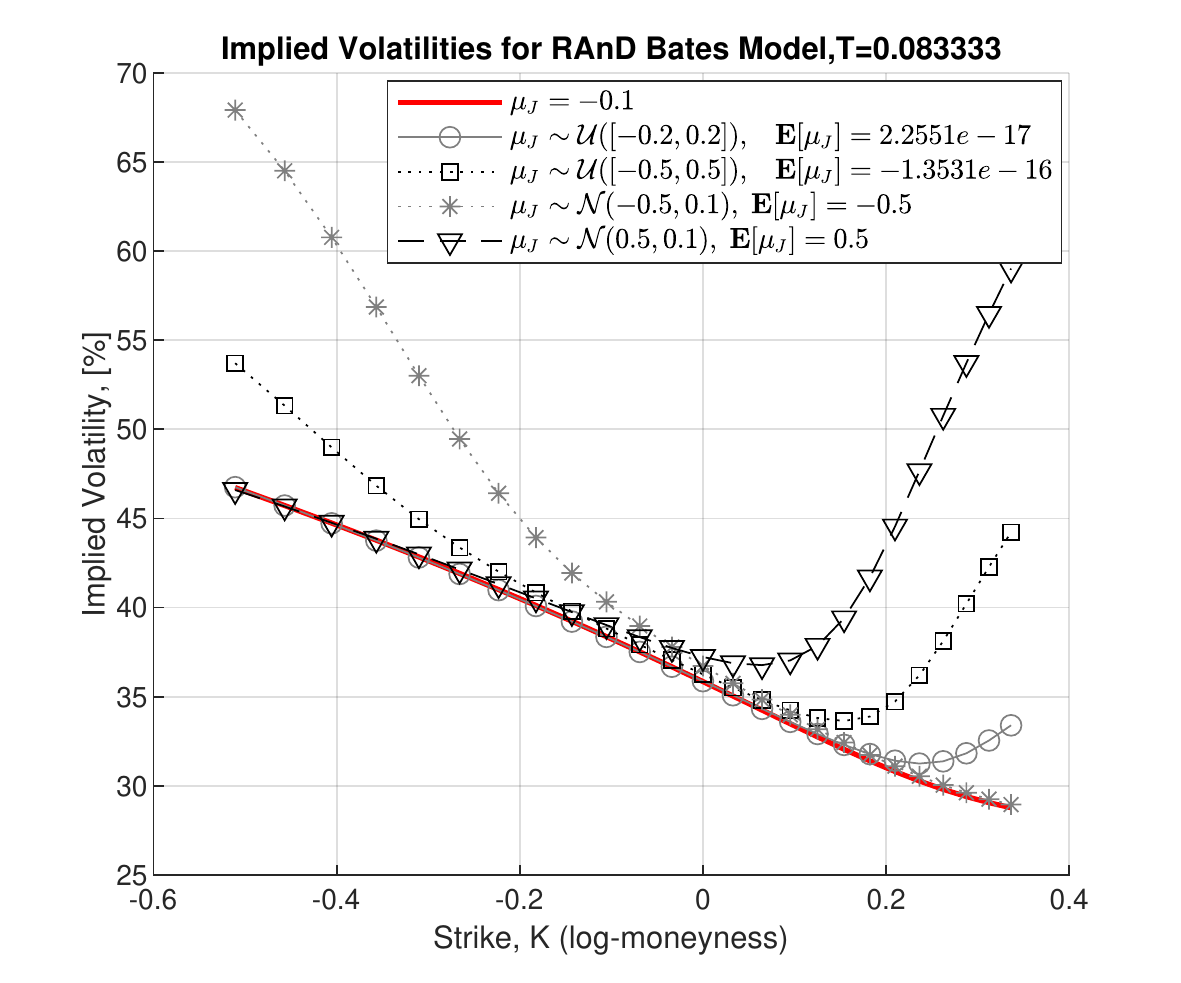}
      \caption{Impact of randomized parameters on implied volatilities. Left: randomized {\it correlation}, $\rho$. Right: randomized {\it jump's mean}, $\mu_J$.}
      \label{fig:impact2}
\end{figure}

Because of the flexibility of the RAnD method, the randomization of all the Bates model parameters can be analyzed. As discussed earlier, one can specify a variety of distributions (see Table~\ref{Tab:Moments}) for the model parameters. In Figure~\ref{fig:impact1} and Figure~\ref{fig:impact2}, the impact of randomization with different distributions and parameters is compared against the benchmark. Randomization enables flexible control of the implied volatility shapes. In this experiment we choose the reference model parameters ($r= 0.0$, $\mu_J = -0.1$, $\sigma_J    = 0.06$, $\lambda = 0.08$, $\kappa     = 0.5$, $\gamma     = 0.5$, $\bar{v}= 0.13$, $\rho = -0.7$,  $v_0= 0.13$ and $T=1/12$), and perform one-by-one randomization. 

Starting with $\gamma$, in the left panel of Figure~\ref{fig:impact1}, we report a significant impact on the curvature and the skew. Randomization of $v_0$ (right panel of Figure~\ref{fig:impact1}) mainly affects the implied volatility level; however, some curvature effect is also present. The randomization of the correlation (left panel of Figure~\ref{fig:impact2}) generates the rotation of the implied volatilities around the ATM level. Another striking effect is reported for $\mu_J$ where only the volatilities far from the ATM level are affected, i.e., one can control either the left or the right tail while keeping the ATM level fixed. The remaining parameters are presented in Figure~\ref{fig:impact3} and Figure~\ref{fig:impact4} in~\ref{sec:appendix_impact}.

%%%%%%%%%%%%%%%%%%%%%%%%%%%%%%%%%%%%%%%%%%%%%%%%%%%%%%%%%%%%%%%%%%%%%%%%%%%%%%%%
\subsection{Vix option pricing under the randomized Heston model with jumps}
\label{sec:vix_Bates}
%%%%%%%%%%%%%%%%%%%%%%%%%%%%%%%%%%%%%%%%%%%%%%%%%%%%%%%%%%%%%%%%%%%%%%%%%%%%%%%%
For a given fixed time-horizon $[t,T],$ the volatility index of an asset $S(t)$, denoted as $\vix(t,T)$ is defined as:
\begin{eqnarray}
\label{eqn:vix_def}
{\overline{\vix}}^2(t,T)=100^2\times\frac{-2}{T-t}\E_t\left[\log \frac{S(T)}{S(t)}\right],
\end{eqnarray}
where $\E_t[\cdot]$ indicates the expectation taken under under the risk-neutral measure $\Q$ and the natural filtration $\F(t)$. 
The VIX formulation in~(\ref{eqn:vix_def}) represents the annualized square root of the price of a contract, where the most liquid options are for $T-t=\text{30 days}.$ The scaling with $100$ relates to a percentage representation of the volatility. 

Depending on the dynamics chosen for stock $S(t)$, the expression in~(\ref{eqn:vix_def}) may be derived in closed-form. In particular, for the case of the Bates model: the closed-form expression for the expectation and the probability density function, $f_{\vix}(x;t, T)$, are known explicitly and are presented in Lemma~\ref{lem:vix}  below.
\begin{lem}
\label{lem:vix}
Under the Bates model~(\ref{eqn:Bates}), the VIX defined in~(\ref{eqn:vix_def}) is expressed by:
\begin{eqnarray}
\label{eqn:Vix_Bates}
\overline\vix^2(t,T)&=&100^2\times\vix^2(t,T),\\
\label{eqn:Vix_Bates2}
    \vix^2(t,T)&=&a(t,T)v(t)+b(t,T)+c,
\end{eqnarray}
where \[a(t,T)=\frac{1-\e^{-\kappa(T-t)}}{\kappa(T-t)},\;\;\;b(t,T)=\bar{v}\left(1-a(t,T)\right),\;\;c=2\lambda\left(\e^{\mu_J+\frac12\sigma_J^2}-\mu_J-1\right),\]
and $v(t)$ follows a non-central chi-square distribution, defined by the CIR process in~(\ref{eqn:Bates}). $\lambda$ represents the Poisson intensity, and $\mu_J$ and variance $\sigma^2_J$ are the parameters for the jumps magnitude, $J\sim\mathcal{N}(\mu_J,\sigma_J^2)$. The PDF of $\vix$ in~(\ref{eqn:Vix_Bates2}), is given by:
\begin{eqnarray}
\label{eqn:vix_chi2}
	f_{\vix}(x;t,T)=2\alpha_1xf_{\chi^2(\delta,\bar \kappa(t_0,t))}\left(\alpha_1(x^2-\alpha_2)\right),
	\end{eqnarray}
where $\alpha_1=\frac{1}{a(t,T)\bar{c}(t_0,t)}$ and $\alpha_2=b(t,T)+c$ with $\delta$, $\bar\kappa(\cdot,\cdot)$ and $\bar{c}(\cdot,\cdot)$ defined in~(\ref{eqn:non-central_params}).
Finally, the ChF for $\vix^2(t,T)$ is given by:
\begin{eqnarray}
\label{eqn:ChF_Vix2}
\phi_{\vix^2(t,T)}(u)=\e^{iu(b(t,T)+c)}\left({\frac{\alpha_1}{\alpha_1-2iu}}\right)^{\frac12\delta}
\exp\left(\frac{iua(t,T)\bar{c}(t_0,t)\bar{\kappa}(t_0,t)}{1-2iua(t,T)\bar{c}(t_0,t)}\right).
\end{eqnarray}
\begin{proof}
The proof can be found in~\ref{appendix:vix}.
\end{proof}
\end{lem}
Given the results from Lemma~\ref{lem:vix} and the availability of the ChF~(\ref{eqn:ChF_Vix2}) for VIX defined in~(\ref{eqn:vix_def}), 
we consider a European option with strike price $K$ and expiry $T$, whose terminal payoff is described as $\max(\vix(T,T+\delta T)-K,0)$, with $\delta T$ is equal to $30$ days. According to
the risk-neutral valuation theory, the option price at time $t,$ denoted by $V_{\vix}(t),$ can be expressed as the discounted conditional expectation of the terminal payoff under the risk-neutral measure, $\Q$:
\begin{eqnarray}
\nonumber
V_{\vix}(t)&=&\e^{-r(T-t)}\E_t\big[\max(\overline{\vix}(T,T+\delta T)-\overline{K},0)\big]\\
&=&100\times\e^{-r(T-t)}\int_{\R^+}\max(\sqrt{v}-{K},0)f_{\vix^2}(v;T,T+\delta T)\d v,\label{eqn:vixOption}
%&=& \e^{-r(T-t_0)}\sideset{}{'}\sum_{k=0}^{N_c-1}
%\Re\left[ \phi_{\bf X}\left(\frac{k\pi}{b-a};t_0,T\right)\exp\left(-i
%k \pi \frac{a}{b-a}\right) \right] \cdot H_k+\epsilon_{c_1},
\end{eqnarray}
for $\overline{K}=100\times K$ and $\overline\vix(\cdot,\cdot)$ defined in~(\ref{eqn:Vix_Bates}).
Since the ChF for $\vix^2(\cdot;\cdot,\cdot)$ is explicitly known (see Equation~(\ref{eqn:ChF_Vix2})), we utilize the Fourier expansion from~(\ref{v3}). The missing ingredients are the payoff-dependent coefficients $H_k$. These coefficients are provided by Lemma~\ref{lem:HkVIX}.
\begin{lem}
\label{lem:HkVIX}
Assuming, $a<{K}^2$, the $H_k$ coefficients, defined in~(\ref{v3}) are given by:
\begin{eqnarray*}
H_k &=&\frac{2}{b-a}\int_a^b\max(\sqrt{y}-{K},0)\cos\left(k\pi \frac{y-a}{b-a}\right)\d y\\
&=:&\frac{2}{b-a}I_1(k)-{K}\frac{2}{b-a}I_2(k),
\end{eqnarray*}
where two integral terms $I_1(k)$ and $I_2(k)$ are given by:
\begin{equation*}
I_1(k):=\left\{
\begin{array}{l l}
	a_0 \Big[\sin ({k_2}) \left(C_F\left(a_1\right)-C_F\left(a_2\right)\right)+\cos ({k_2}) \left(S_F\left(a_2\right)-S_F\left(a_1\right)\right)\Big]\\
+a_5 \Big[\sqrt{b}\sin (a_3)-{K} \sin \left(a_4\right)\Big],&k\neq0,\\
&\\
\frac{2}{3}(b^{\frac{3}{2}}-{K}^3),& k=0,\qquad
\end{array}
\right.\end{equation*}
and
\begin{equation*}
I_2(k):=\left\{
\begin{array}{l l}
	a_5\left[\displaystyle\sin\left(a_3\right) + \sin\left(-a_4\right)\right],&k\neq0,\\
&\\
(b-K^2),& k=0,\qquad
\end{array}
\right.\end{equation*}
with $k_1=\frac{k\pi}{b-a}$, $k_2=\frac{ak\pi}{b-a}$, $a_0=\frac{\sqrt{\frac{\pi }{2}}}{{k_1}^{3/2}}$, $a_1=\sqrt{bk_1\frac{2}{\pi }},$ $a_2={K} \sqrt{{k_1}\frac{2}{\pi }}$, $a_3=b {k_1}-{k_2}$, $a_4={K}^2 {k_1}-{k_2}$, $a_5=\frac{1}{{k_1}}$ and where $C_F(\cdot)$ and $S_F(\cdot)$ are the so-called Fresnel integrals defined as:
\[C_F(x)=\int_0^x\cos(t^2)\dt,\;\;\;S_F(x)=\int_0^x\sin(t^2)\dt.\]
\begin{proof}
After the change of variables, $x=\sqrt{y},$ the integration is straightforward.
\end{proof}
\end{lem}
By semi-analytic expressions for pricing options on the underlying $S(t)$, via Equation~(\ref{v3}) and on the VIX, in~(\ref{eqn:vixOption}), we can calibrate the randomized AD models. The pricing formulae depend on the integration ranges $[a,b]$ that needs to be chosen carefully. It is advised to choose it based on the cumulants, which have been derived for VIX derivatives in~\cite{jing2021consistent}. We report, however, that calibration of VIX derivatives may result in rather extreme parameters giving rise to numerical instabilities. Since we can utilize the analytically known distribution for $\overline{\vix}$, the pricing may also be performed by directly integrating the payoff function and employing the PDF in~(\ref{eqn:vix_chi2}):
\begin{eqnarray*}
V_{\vix}(t)=100\times 2\alpha_1\e^{-r(T-t)}\int_{\overline{K}}x(x-\overline{K})f_{\chi^2(\delta,\bar \kappa(t,T))}\left(\alpha_1(x^2-\alpha_2)\right)\dx,
\end{eqnarray*}
with $\alpha_1$ and $\alpha_2$ in~(\ref{eqn:vix_chi2}) and $\delta$, $\bar\kappa(t,T)$ defined in~(\ref{eqn:non-central_params}). With one of the model parameters stochastic, the RAnD pricing equation reads:
\begin{eqnarray*}
V_{\vix}(t)%&=&100\times 2\alpha_1\e^{-r(T-t)}\int_{\overline{K}}x(x-\overline{K})f_{\chi^2(\delta,\bar \kappa(t,T))}\left(\alpha_1(x^2-\alpha_2)\right)\dx%\\
=100\times 2\alpha_1\e^{-r(T-t)}\sum_{n=1}^N\omega_n\int_{\overline{K}}x(x-\overline{K})f_{\chi^2(\delta,\bar \kappa(t,T))}\left(\alpha_1(x^2-\alpha_2);\theta_n\right)\dx,
\end{eqnarray*}
where $\theta_n$ in $f_{\chi^2(\cdot,\cdot)}\left(\cdot;\theta_n\right)$ indicates a particular realization of the model parameter and $\omega_n$ corresponds to its weight.

%%%%%%%%%%%%%%%%%%%%%%%%%%%%%%%%%%%%%%%%%%%%%%%%%%%%%%%%%%%%%%%%%%%%%%%%%%%
\subsection{Market data experiment: calibration to S\&P500 and VIX options}
\label{sec:calibrationVIX}
%%%%%%%%%%%%%%%%%%%%%%%%%%%%%%%%%%%%%%%%%%%%%%%%%%%%%%%%%%%%%%%%%%%%%%%%%%%
This section illustrates how the randomized Bates (RAnD Bates) model can fit implied volatilities for S\&P 500 and VIX. In the experiment, we consider three randomly selected dates: 02/02/2022, 13/05/2022 and 14/07/2022, for which the calibration exercise was performed. In addition, we have considered short expiry options of one month in all the cases- a challenging task for conventional stochastic volatility models. The calibration was performed in two stages; in the first step, the {\it pure} Bates model was calibrated to European S\&P 500 options, and such attained parameters form an initial guess for the RAnD Bates model. Then, the optimization was performed with the {\it target} function, including implied volatilities on S\&P 500 and VIX. In the calibration, an additional weight for the ATM level was used. In this calibration stage, we have used the randomized vol-vol parameter $\gamma$ and randomized it with a uniformly distributed random variable, $\gamma\sim\mathcal{U}([\hat a,\hat b])$. 

\begin{figure}[h!]
  \centering
    \includegraphics[width=0.45\textwidth]{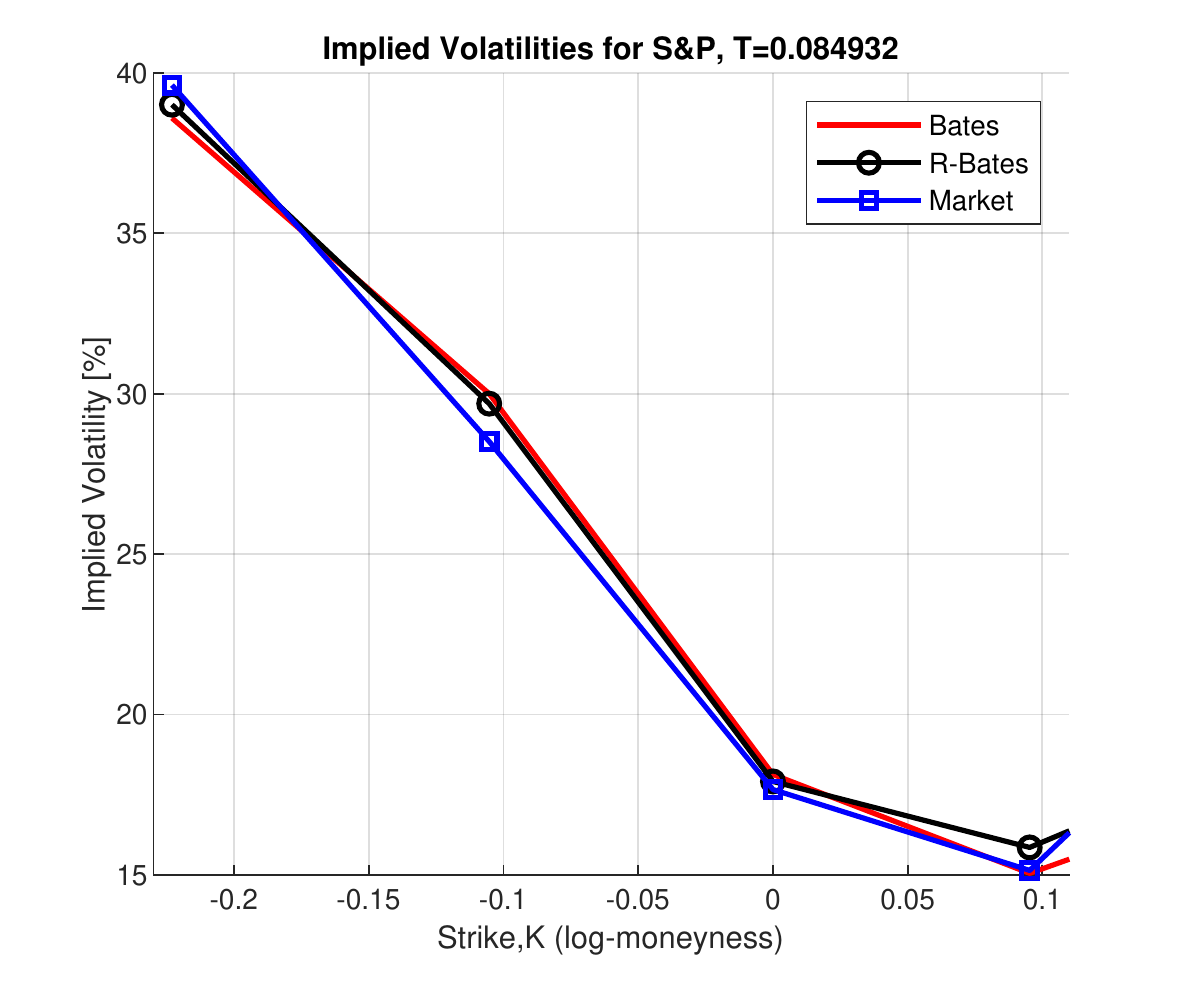}
    \includegraphics[width=0.45\textwidth]{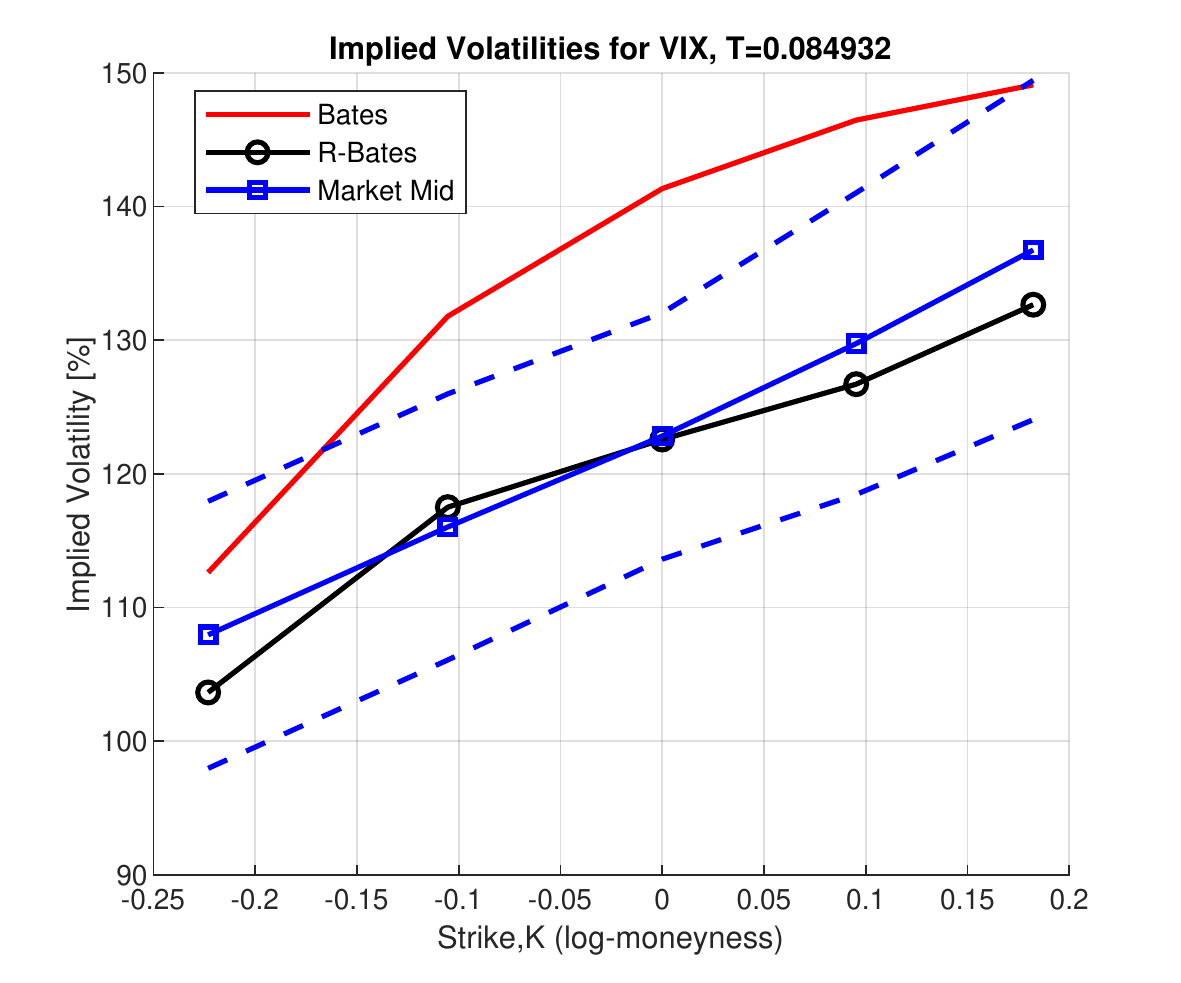}
      \caption{Calibration results of the RAnD Bates model. The implied volatilities for S\&P and ViX were obtained on  02/02/2022. Dotted lines indicate bid-ask spreads. Left: S\&P, Right: VIX. }
      \label{fig:SPX_VIX1}
\end{figure}

\begin{figure}[h!]
  \centering
    \includegraphics[width=0.45\textwidth]{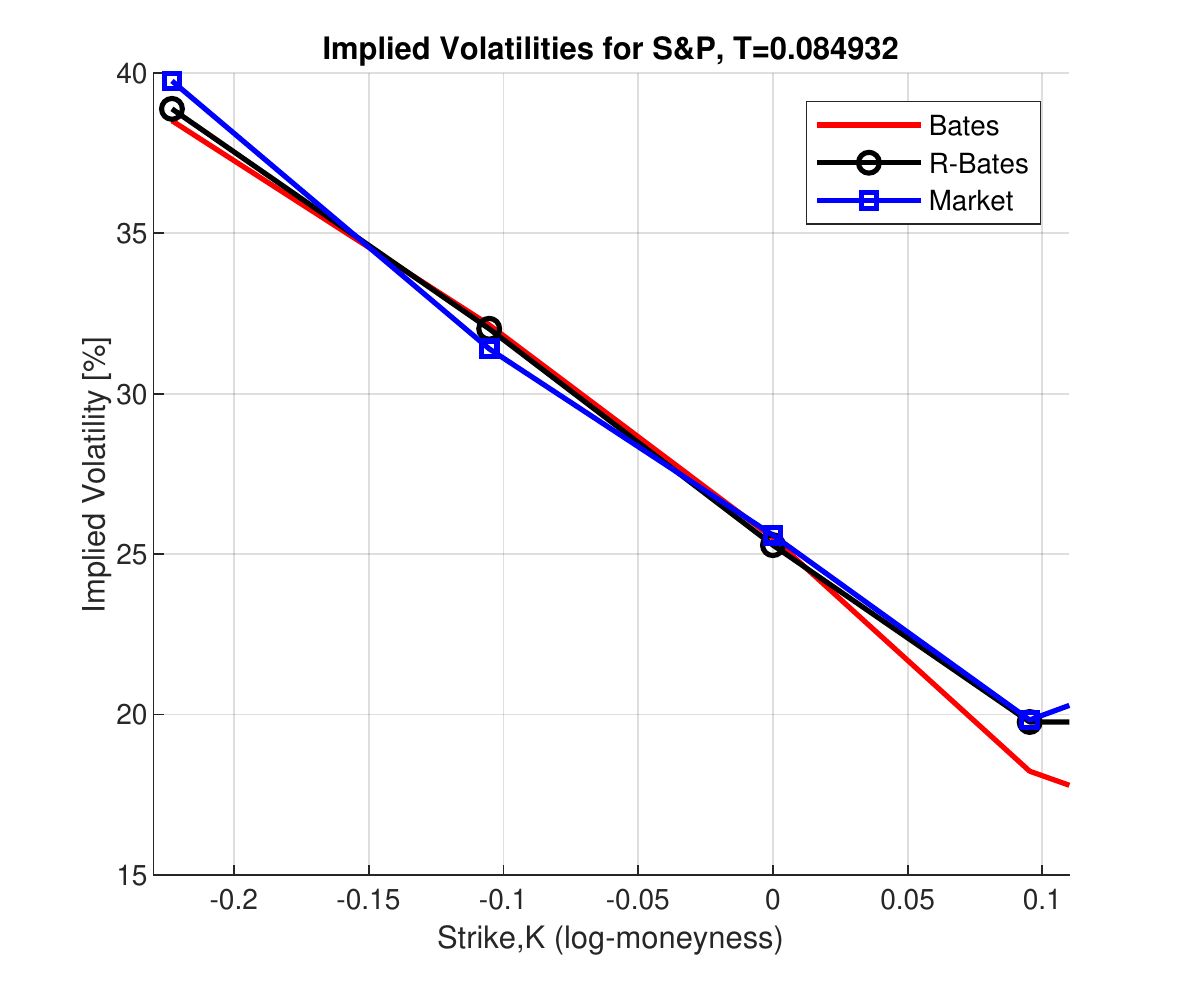}
    \includegraphics[width=0.45\textwidth]{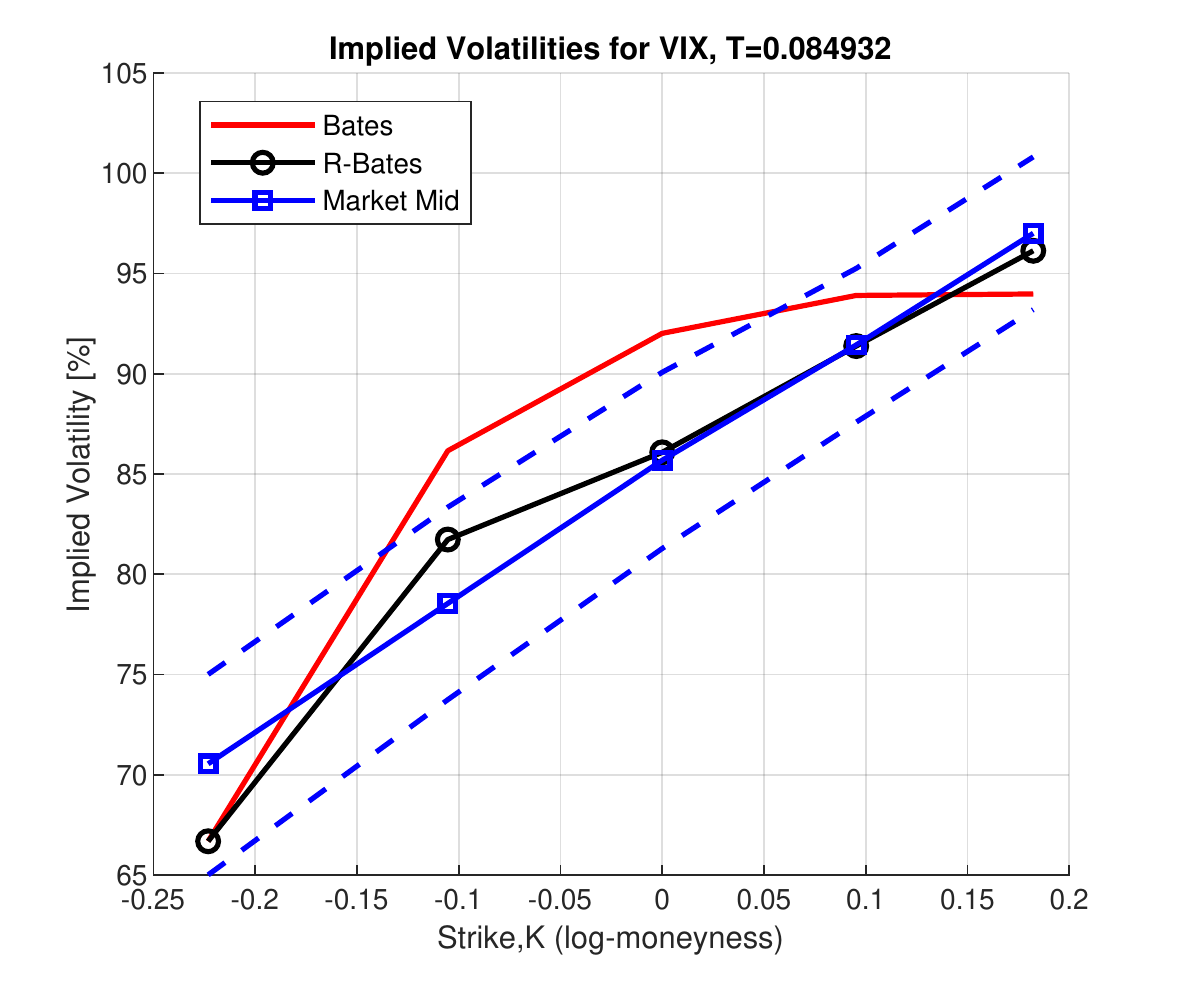}
       \caption{Calibration results of the RAnD Bates model. The implied volatilities for S\&P and ViX were obtained on  14/07/2022. Dotted lines indicate bid-ask spreads. Left: S\&P, Right: VIX. }
      \label{fig:SPX_VIX2}
\end{figure}

\begin{figure}[h!]
  \centering
    \includegraphics[width=0.45\textwidth]{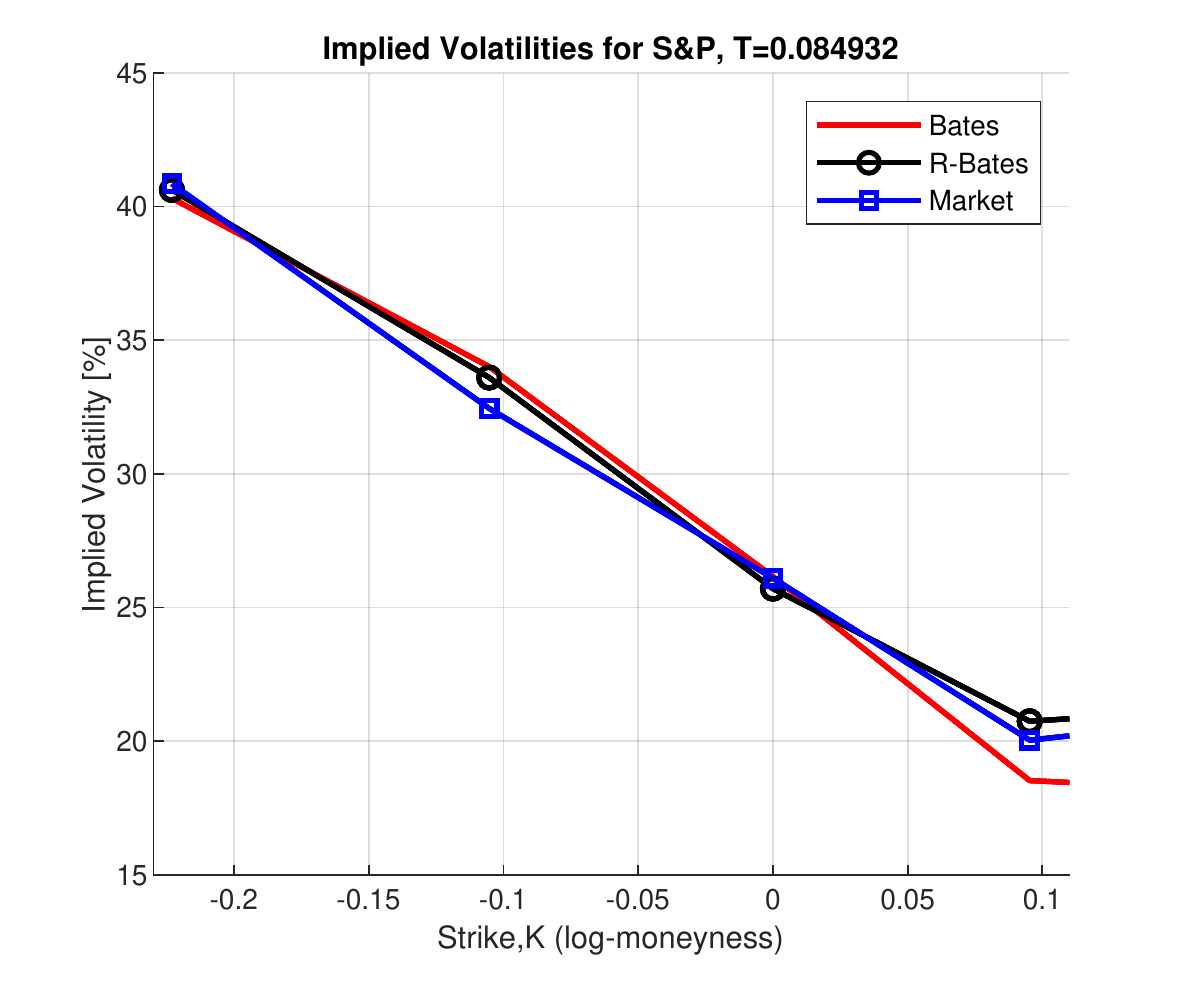}
    \includegraphics[width=0.45\textwidth]{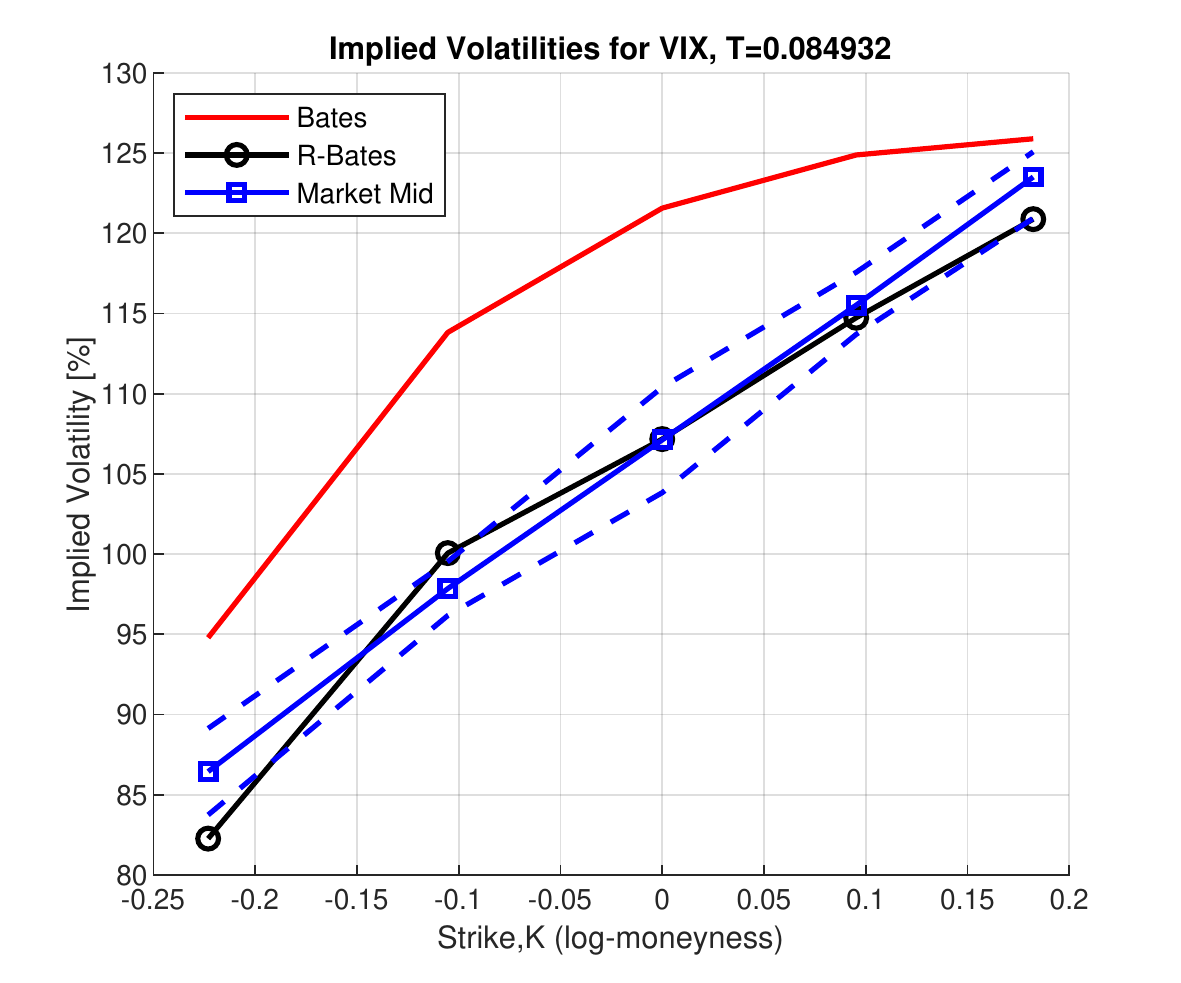}
      \caption{Calibration results of the RAnD Bates model. The implied volatilities for S\&P and ViX were obtained on  13/05/2022. Dotted lines indicate bid-ask spreads. Left: S\&P, Right: VIX. }
      \label{fig:SPX_VIX3}
\end{figure}

The calibration results are presented in Figures~\ref{fig:SPX_VIX1},\ref{fig:SPX_VIX2},\ref{fig:SPX_VIX3}. For all three cases, the results were compared against the Bates model (indicated by the red line). We conclude that jumps in the Bates model allow for a good fit for European options on the index (illustrated in the left panels). However, it is also clear that the Bates model does not have sufficient flexibility to fit VIX options (right panels). The standard Bates model gives results significantly outside the bid-ask spreads for all three studied cases. 

The calibrated parameters for the RAnD Bates model are reported in Table~\ref{Tab:CalibrationBates}. The results are excellent, i.e., for all three cases, we can calibrate to both products simultaneously, and the implied volatilities for the RAnD Bates model for VIX were well within the bid-ask spreads.

\begin{table}[htb!]
\centering\footnotesize
\caption{\footnotesize Parameters determined in calibration of S\&P and VIX}
\begin{tabular}{c|c|c|c|c|c|c|c|c}
\multicolumn{8}{c}{$\text{Calibrated RAnD Bates parameters}$}\\
date&$\kappa$&$v_0$&$\bar{v}$&$\rho$&$\mu_J$&$\sigma_J$&$\lambda$&$\gamma$\\\hline\hline
02/02/2022&0.5&$0.170^2$&0.23&-0.65&$-0.25$&0.05& 0.25&	$\gamma\sim\mathcal{U}([0.01,2.3])$\\
13/05/2022&0.14&$0.267^2$&0.28&-0.8&$-0.25$&0.02& 0.1&$\gamma\sim\mathcal{U}([0.002,2.1])$\\
14/07/2022&0.5&$0.250^2$&0.10& -0.85&$-0.25$&0.05&0.15&$\gamma\sim\mathcal{U}([0.05,1.4])$\\
\end{tabular}
\label{Tab:CalibrationBates}
\end{table}

It is important to remember that the calibration of the randomized models, especially for models like the Bates model with multiple parameters for stochastic volatility and jumps, is not trivial. For that reason, we have used the standard Bates model parameters as the initial guess for the RAnD framework- otherwise, one may encounter the well-known problem of the possibility of ending up in a local minimum. 

%%%%%%%%%%%%%%%%%%%%%%%%%%%%%%%%%%%%%%%%%%%%%%%%%%%%%%%%%%%%%%%%%%%%%%%%%%%
\subsection{Hedging under the RAnD method }
\label{sec:hedging}
%%%%%%%%%%%%%%%%%%%%%%%%%%%%%%%%%%%%%%%%%%%%%%%%%%%%%%%%%%%%%%%%%%%%%%%%%%%
Using the pricing Equation~(\ref{v3}), one can find the hedge parameters $\Delta$ and $\Gamma$ explicitly~\citep{OosterleeGrzelakBook}. However, it is a rather involved task to compute the sensitivity to the model parameters, or the parameters of the randomizer need to be computed. 

Typically, when the market data changes, a model needs to be re-calibrated, requiring a trader to {\it hedge} each parameter. Applying the RAnD method allowed us to make use of randomized processes; it led us to a marginal density for the asset price as a convex combination of affine densities (see Remark~\ref{rem:RandDensity}). Conversely, the option price can be represented as the same convex combination among the corresponding option prices. Moreover, due to the linearity of the derivative operator, the same convex combination applies to all option Greeks. In particular, this ensures that starting from analytically tractable affine densities, one deals with a model that preserves the analytical tractability. 
Let us consider a scenario in which a model parameter $\theta$ follows a specific distribution with a parameter $\hat a$. Under such a randomized model, the sensitivity of the derivative price, $V(t_0)$, should be calculated to the parameter, $\hat a$, that drives the parameter's randomness. The RAnD framework uses the randomized ChF, which must be differentiated to compute the sensitivities. By differentiation of the pricing, Equation~(\ref{v3}), for 
$\bar{A}:=\bar A({ u};\tau,\theta_n)$ and $\bar{B}:={\bf{\bar B^\T}}({ u};\tau,\theta_n)$, we find:
\begin{equation}\label{eqn:sensies}
\frac{\partial V(t_0)}{\partial \hat a}= \e^{-r(T-t_0)}\sideset{}{'}\sum_{k=0}^{N_c-1}
\Re\left[\exp\left(-i
k \pi \frac{a}{b-a}\right) \frac{\partial}{\partial \hat a}\phi_{\bf X}\left(\frac{k\pi}{b-a};t_0,T\right)\right] \cdot H_k+\epsilon_{c_2},
\end{equation}
with:
\begin{eqnarray}
\label{eqn:sensies2}
\frac{\partial}{\partial \hat a}\phi_{\bf X}\left(u;t_0,T\right)=\sum_{n=1}^{N}\phi_{{\bf X}|\vartheta={\theta_n}}(u)\left[\frac{\partial \omega_n}{\partial \hat a}+\omega_n\left(\frac{\partial \bar{A}}{\partial \theta_n}+{\bf X}(t)\frac{\partial \bar B}{\partial \theta_n}\right)\frac{\partial \theta_n}{\partial \hat a}\right],
\end{eqnarray} where $\phi_{{\bf X}|\vartheta={\theta_n}}(u)=\e^{\bar A({\bf u};\tau,\theta_n)+{\bf{\bar B^\T}}({\bf u};\tau,\theta_n){\bf{X}}(t)}$, with $H_k$ the payoff coefficient (independent of the model parameters). Error $\epsilon_{c_1}$ is a function of $\epsilon_{c_1}$ in~(\ref{v3}) and the quadrature error, $\epsilon_N$, in~(\ref{eqn:randChF}).
The expression in~(\ref{eqn:sensies2})  involves derivatives of the ChF with respect to the model parameters $\partial \bar{A}/\partial \theta_n$ and $\partial \bar{B}/\partial \theta_n$,  which may to be derived analytically, and the sensitivity to the quadrature pairs, $\{\omega_n,\theta_n\}$, $\partial \theta_n/\partial \hat a$ and $\partial \omega_n/\partial \hat a$. Since the computation of these pairs is not analytic, it is recommended to compute these derivatives numerically, with, for example, finite differences:
\[\frac{\zeta(\vartheta(\hat a+\delta_{\hat a}))-\zeta(\vartheta(\hat a-\delta_{\hat a}))}{2\delta_{\hat a}}\approx \left\{\frac{\partial \omega_n}{\partial \hat a},\frac{\partial \theta_n}{\partial \hat a}\right\},\]
where $\vartheta(\hat a)$ indicates the dependence of the random variable and parameter $\hat a$, $\delta_{\hat a}$ is the {\it shock} size and $\zeta(\vartheta):\R\rightarrow \{\omega_n,\theta_n\}_{n=1}^N$ is defined in~\ref{app:collocationTheory}. Due to the applied finite difference shocks to $\hat a$, an additional bias will be introduced. We expect, however, this error to be of acceptable magnitude. 

%%%%%%%%%%%%%%%%%%%%%%%%%%%%%%%%%%%%%%%%%%%%%%%%%%%%%%%%%%%%%%%%%%
\subsection{Convergence improvements for Monte Carlo simulation with randomized parameters}
\label{sec:convergence}
%%%%%%%%%%%%%%%%%%%%%%%%%%%%%%%%%%%%%%%%%%%%%%%%%%%%%%%%%%%%%%%
Under the RAnD framework, the randomness of the model parameters is not modelled by an AD SDE, but it is enforced ``externally'' and consequently affects the simulation with Monte Carlo methods and the convergence. In this section, we discuss how to perform Monte Carlo sampling and propose an efficient variance reduction technique that will improve the convergence of the method. 

Monte Carlo simulation of the randomized model implies that every simulated path of the underlying process ${\bf X}(t)$ will be randomized in a sense that will depend on a realization of $\vartheta$, $\theta_n$. 
Suppose we consider a discretization of some process ${\bf X}(t)$ from~(\ref{ajd:ch1}) with $M$ Monte Carlo paths given by:
\begin{eqnarray}
\nonumber
\small
	n: \;{{x}}_n(t+\Delta t;\theta_n)&=&{{x}}_n(t;\theta_n)+\int_t^{t+\Delta t}\bar{\mu}(s,{{x}}_n(s);\theta_n)\ds+\int_t^{t+\Delta t}\bar{\sigma}(s,{\bf{x}}_n(s);\theta_n)\d{\widetilde{{w}}_n}(s)\\&+&\int_t^{t+\Delta t}{ J}_n(s;\theta_n)^\T\d { x}_{\mathcal{P},n}(s),
	\label{eqn:McPath}
\end{eqnarray}
with ${\widetilde{{w}}_n}(s)$ and ${ x}_{\mathcal{P},n}(s)$ representing the realization of the Brownian motion and Poisson process, respectively. Then, depending on the discretization method, the integrals above can be found explicitly or approximately by either Euler, Milstein or any other higher-order discretization scheme. Since the randomization of model parameters is time-invariant, this implies that every discretized Monte Carlo path of~(\ref{eqn:McPath}) will be driven by a different realization of the randomized parameter. Formally, considering a simulation with $M$ paths, every $n$'th, $n=1,\dots,M$,  path will depend on the $n$'th sample from the parameter distribution,

The computational time can be reduced by following a similar strategy as in Section~\ref{sec:RAnD_BS}. We have two possibilities for derivative pricing under the RAnD framework: directly using all the Monte Carlo paths and evaluate the pricing equation, $V_1(t_0)$ in~(\ref{eqn:pricingEqn_MC}), or we can employ the idea behind the RAnD method and divide the pricing problem into $N$ sub-problems, where for each quadrature node $\{\theta_n\}_{n=1}^N$, a separate simulation is performed and weighted according to $\{\omega_n\}_{n=1}^N$ (see $V_2(t_0)$ in Equation~(\ref{eqn:pricingEqn_MC})),
\begin{eqnarray}
\label{eqn:pricingEqn_MC}
V_1(t_0)\approx\frac{\e^{-rT}}{M}\sum_{i=1}^M\Pi({ x}_i(T),T),\;\;\;V_2(t_0)\approx\frac{\e^{-rT}}{M}\sum_{n=1}^{N}\omega_n \sum_{i=1}^M\Pi({ x}_i(T;\vartheta=\theta_n),T).
\end{eqnarray}

The division into sub-problems gives stability to the model; however, the cost is the quadrature error, affecting the overall pricing error. 

\begin{figure}[h!]
  \centering
    \includegraphics[width=0.45\textwidth]{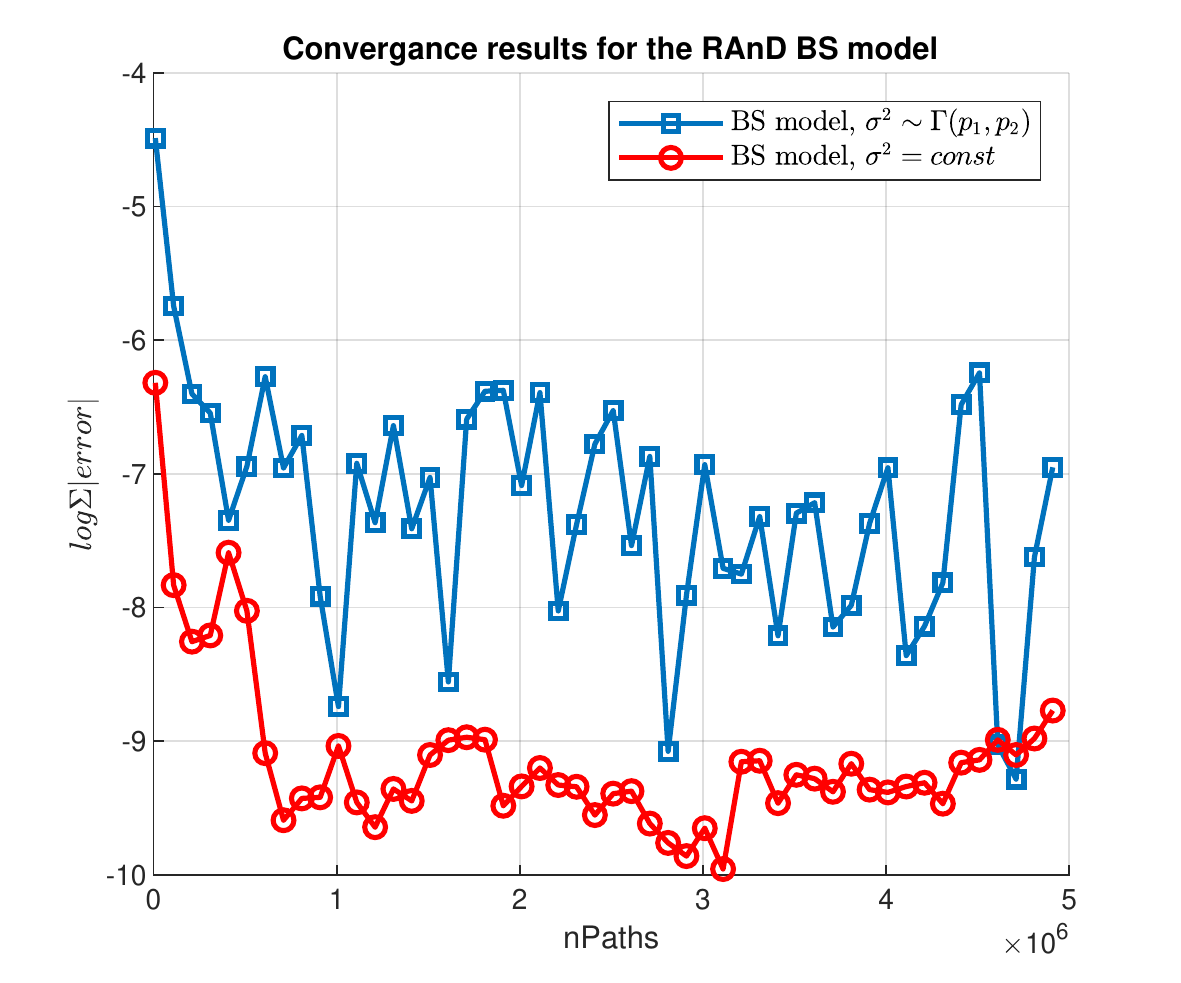}
    \includegraphics[width=0.45\textwidth]{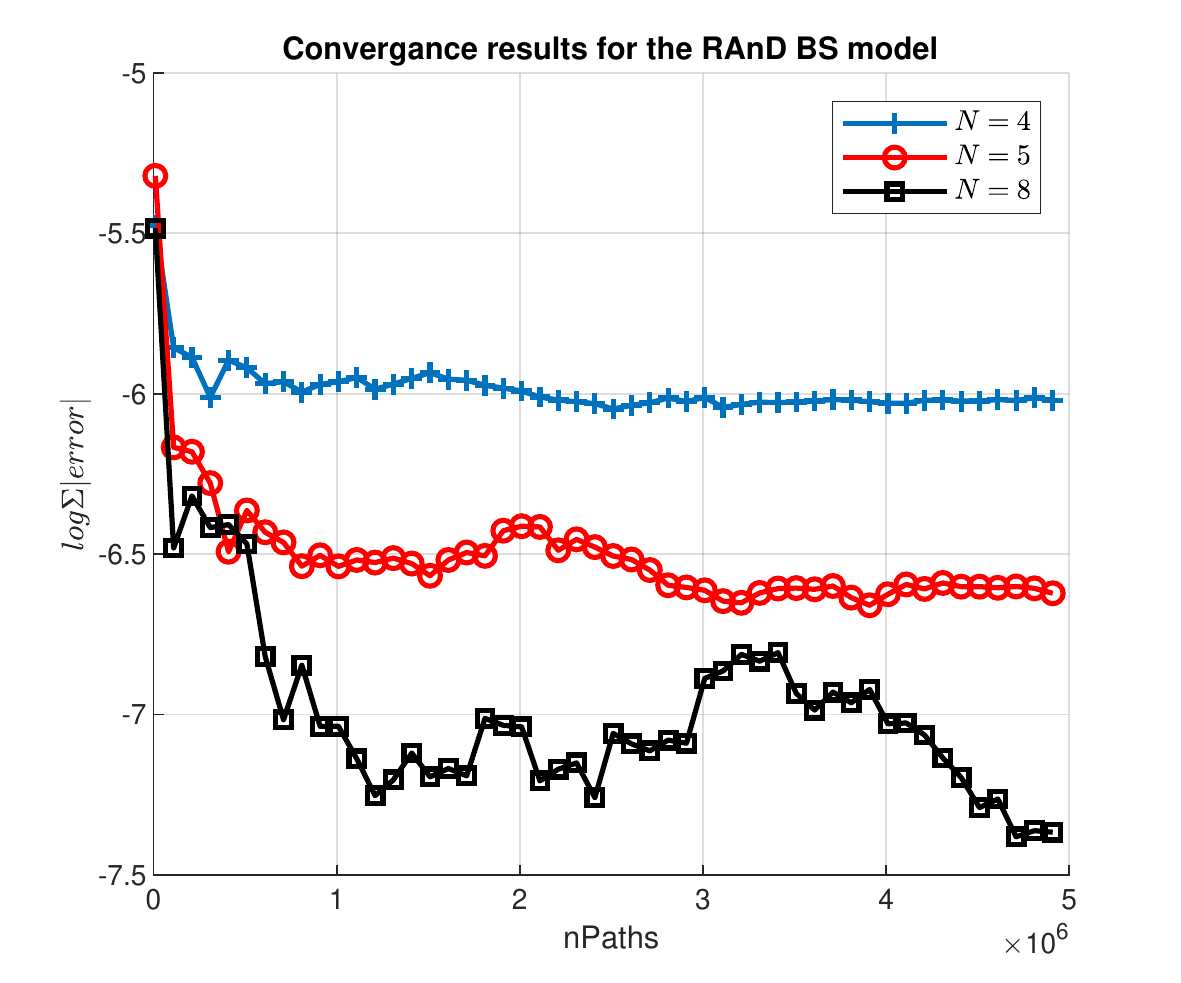}
      \caption{Convergence results for an increasing number of Monte Carlo paths. Left panel: impact of randomization on convergence rate. Right panel: RAnD method convergence with sub-divided pricing.}
      \label{fig:convegence_Paths}
\end{figure}

In Figure~\ref{fig:convegence_Paths}, the numerical results for the example case discussed in Section~\ref{sec:RAnD_BS} are presented. The left panel of Figure~\ref{fig:convegence_Paths} confirms that the stability is affected by the randomized volatility coefficient. In the right panel of Figure~\ref{fig:convegence_Paths}, we see stability by dividing the problem into $N$ sub-problems. Although the division process improves the convergence, it will negatively impact the computational cost (the pricing needs to be performed $N$ times). Finally, a particular division strategy will also depend on the pricing problem.

%%%%%%%%%%%%%%%%%%%%%%%%%%%%%%%%%%%%%%%%%%%%%%%%%%%%%%%%%%%%%%%%%%%%%%
\subsection{RAnD method for piece-wise constant parameters}
\label{sec:piece_wise}
%%%%%%%%%%%%%%%%%%%%%%%%%%%%%%%%%%%%%%%%%%%%%%%%%%%%%%%%%%%%%%%%%%%%%%
The methodology to obtain the randomized ChF can be extended to multiple or piece-wise parameters. In such a case, the approach could be considered a rough approximation for a parameter driven by a stochastic process. Using the insight that under the affine models functions $\bar A(u,\tau)$ and $\bar{\bf B}(u;\tau)$ in~(\ref{eqn:randChF}) are the solution of the Riccati-type ODEs with nonzero initial conditions $\bar A(u,0)$ and $\bar {\bf B}(u,0)$, for a time-grid $\tau_{n_1}\leq \tau_{n_2}\dots\leq \tau_{n_m}=\tau$, the model parameters can be evaluated at each interval determined by $T-\tau_i$. Piece-wise constant parameters imply that the characteristic function can be evaluated recursively, i.e. at the first interval from
$[0,\tau_{n_1})$, we use the initial conditions $\bar{{\bf B}}({ u},0)=0$ and
$\bar{A}({u},0)=0$. When the corresponding analytic solution is determined (for details see~\citep{OosterleeGrzelakBook}, p.521), we obtain two solutions, $a_1$ and $c_1$. For
interval $[\tau_1,\tau_2)$, we then assign the initial conditions
$\bar{{\bf B}}({ u},\tau_1)=c_1$ and $\bar{A}({ u},\tau_1)=a_1$. This procedure will
be repeated until the last time step, where the initial values $c_{n_m-1}$ and
$a_{n_m-1}$ are used to evaluate $\bar{{\bf B}}({ u},\tau_{n_m})$ and $\bar{A}({u},\tau_{n_m})$.

Piece-wise parameters divided over $n_m$ intervals will require separate handling of each interval. Corollary~\ref{cor:piecewise} provides the details.
\begin{cor}[Piece-wise random parameters] Consider $n_m$ independent random variables $\vartheta_1,\dots,\vartheta_{n_m}$, and a time grid with $n_m$ intervals of the time interval $[t,T]$ with $\tau=T-t$. Each of the variables corresponds to an interval. Then, the corresponding randomized ChF is given by:
\label{cor:piecewise}
\begin{eqnarray}
\label{eqn:piecewise}
\phi_{{\bf X}}({ u};t,T)=\sum_{n_1=1}^{N}\dots\sum_{n_m=1}^{N}\omega_{n_1}\dots\omega_{n_{m}}\e^{\bar A({ u};\tau,\theta_{n_1},\dots,\theta_{n_m})+{\bf{\bar B^\T}}({ u};\tau,\theta_{n_1},\dots,\theta_{n_m}){\bf{X}}(t)}+\bar\epsilon_N,
\end{eqnarray}
under the same assumption as specified in Theorem~\ref{prop:RAnDChF}.
\end{cor}
The representation above gives us an extension for $n_m$ independent model parameters. The summation is over all parameters and their corresponding quadrature points; thus, it is subject to the so-called {\it curse of dimensionality}. Therefore, this expression is not desirable for many stochastic parameters. However, the representation in~(\ref{eqn:piecewise}) can be reduced using the sparse grid approach (see~\cite{Smol63,grzelak2021sparse}). The significant advantage of such a technique is that the number of grid points does not grow exponentially with the dimension but only polynomially.

%%%%%%%%%%%%%%%%%%%%%%%%%%%%%%%%%%%%%%%%%%%%%%%%%%%%%%%%%%%%%%%
\section{Conclusion}
\label{sec:conclusion}
%%%%%%%%%%%%%%%%%%%%%%%%%%%%%%%%%%%%%%%%%%%%%%%%%%%%%%%%%%%%%%%
In this article, we have introduced the RAnD method for efficient computation of the affine models with random parameters. The proposed framework is generic and can be applied to any stochastic model, even outside the class of affine diffusions. As long as the randomizing random variable gives rise to finite, preferably closed-form, moments, one can price European-style options efficiently. The heart of the method is formed by a few {\it critical} collocation points to recover the characteristic function. Fast computation of the characteristic function is possible because the method converges exponentially in the number of expansion terms. We have shown that the randomization of stochastic models provides a breeze of fresh air to the class of affine models. Finally, we have applied the RAnD method to the Bates model and shown that randomization allows for simultaneous calibration to S\&P and VIX options- a heavily desired feature of modern models.

\bibliographystyle{abbrv}
\bibliography{bibliography.bib}

\appendix 
%%%%%%%%%%%%%%%%%%%%%%%%%%%%%%%%%%%%%%%%%%%%%%%%%%%%%%
\section{Additional materials and proofs}
%%%%%%%%%%%%%%%%%%%%%%%%%%%%%%%%%%%%%%%%%%%%%%%%%%%%%%

%%%%%%%%%%%%%%%%%%%%%%%%%%%%%%%%%%%%%%%%%%%%%%%%%%%%%%%%%%%%%
\subsection{\bf Moments and optimal collocation pairs: $\zeta(\vartheta):\R\rightarrow \{\omega_n,\theta_n\}_{n=1}^N$}
\label{app:collocationTheory}
%%%%%%%%%%%%%%%%%%%%%%%%%%%%%%%%%%%%%%%%%%%%%%%%%%%%%%
%In this section we discuss how to determine optimal collocation points $x_i$.

This section discusses the mechanism for the construction of orthogonal polynomials. Later, this methodology is applied to establish the $\zeta(\vartheta):\R\rightarrow \{\omega_n,\theta_n\}_{n=1}^N$ that are used in the RAnD method.

A sequence of orthogonal polynomials $\{p_i\}_{i=0}^N$, with
$\deg(p_i)=i$, is said to be orthogonal in $L^2$ with respect to PDF
$f_\vartheta(\vartheta)$ of $\vartheta$, if the following holds,
\begin{equation}\label{eqPsi}
\E\left[p_i(\vartheta)p_j(\vartheta)\right] = \int_\R p_i(x) p_j(x)f_\vartheta(x)\d x =
\delta_{i,j}\E\left[p_i^2(\vartheta)\right],\;\;\;i,j=0,\dots,N,
\end{equation}
with $\R$ the support of $\vartheta$, $\delta_{i,j}$ the Kronecker delta.

An important property of orthogonal polynomials is their definition in terms of a recurrence relation,
given in the theorem below.
\begin{thm}[Recurrence in orthogonal polynomials]
\label{thm:recurrence} For any given density function $f_\vartheta(\cdot)$,
a unique sequence of monic orthogonal polynomials ${p}_i(x)$ exists,
with $\text{deg}({p}_i(x))=i$, which can be constructed as follows,
\begin{eqnarray}
\label{psi:reccurence}
{p}_{i+1}(x)=(x-\alpha_i){p}_i(x)-\beta_i{p}_{i-1},\;\;\;i=0,\dots,N-1,
\end{eqnarray}
where ${p}_{-1}(x)\equiv 0$, ${p}_0(x)\equiv 1$ and where $\alpha_i$
and $\beta_i$ is the recurrence coefficients,
\begin{eqnarray}
\label{4.10}
\alpha_i=\frac{\E[\vartheta{p}_i^2(\vartheta)]}{\E[{p}_i^2(\vartheta)]},\;\mbox{for }\;i=0,\dots,N-1,\;\;\;\;\beta_i=\frac{\E[{p}_i^2(\vartheta)]}{\E[{p}_{i-1}^2(\vartheta)]},\;\mbox{for }\;i=1,\dots,N-1,
\end{eqnarray}
with $\beta_0=0$.
\begin{proof} The proof can be found in~\citep{Favard:OrthogonalPoynomials}.
\end{proof}
\end{thm}

Parameters $\alpha_i$ and $\beta_i$ are entirely determined in terms of the moments of a random variable $\vartheta$.  For many densities (weight functions in the integration in~(\ref{eqPsi})) the three-term recurrence relation in~(\ref{psi:reccurence}) of the corresponding orthogonal polynomials has been determined. In a scenario when the probability density function $f_\vartheta(\cdot)$ is not known explicitly, or its evaluation is computationally expensive.  In such circumstances, it is desirable to express $\alpha_i$ and $\beta_i$ in~(\ref{4.10}) in terms of the moments of $\vartheta$~\citep{GoWe}.  Let us consider the monomials $m_i(\vartheta)=\vartheta^i$, and define $\mu_{i,j}$ as follows,
\begin{equation}\label{eqPsi}
\mu_{i,j}=\E\left[{m}_i(\vartheta){m}_j(\vartheta)\right] =
\int_\R x^{i+j}f_\vartheta(x)\d x =\E[\vartheta^{i+j}],\;\;\;i,j=0,\dots,N.
\end{equation}
From all moments $\mu_{i,j}$ we construct the so-called Gram matrix
$M=\{\mu_{i,j}\}_{i,j=0}^N$, which is symmetric and contains all moments $\{1,\E[\vartheta^1],\dots,\E[\vartheta^{2N}]\}$. Since matrix $M$ is, by
definition, positive definite~\citep{GoWe}, we decompose
$M=R^\T R$, by the Cholesky decomposition of $M.$

The following step relates the Cholesky upper triangular matrix $R$ to the orthogonal polynomials.  This relationship has been established
in~\citep{GoWe} and is given by,
\begin{eqnarray}
\label{eqn:alphaBeta}
\alpha_j=\frac{r_{j,j+1}}{r_{j,j}}-\frac{r_{j-1,j}}{r_{j-1,j-1}},\;\;\; j=1,\dots,N,\;\;\;\text{and}\;\;\;\beta_j=\left(\frac{r_{j+1,j+1}}{r_{j,j}}\right)^2,\;\;\;j=1,\dots,N-1,
\end{eqnarray}
with $r_{0,0}=1$ and $r_{0,1}=0$ and where $r_{i,j}$ is the $(i,j)$-th element of matrix $R$. This relation gives us the manifestations for
$\alpha_j$ and $\beta_j$ when the matrix of moments has been computed.

%\begin{algorithm}
%\caption{Calculation of coefficients $\bff \alpha$ and $\bff \beta$ for
%the orth. polynomials $\bff p(X).$} \label{alg:1} \small {\text
%// Construction of the matrix $M$:}\\\For{$i,j=1\ldots N+1$}{
% $M[i,j]=\E[X^{i+j-2}]$ } // Calculate the Cholesky matrix from $M$:\\
%\KwResult{$M=RR^\T$} // Find $\bff \alpha$ and $\bff \beta$ from matrix
%$R$\\
%$\alpha[1]=R[1,2]/R[1,1]$\\
%$\beta[1]=(R[2,2]/R[1,1])^2$\\
%\For{$i=2:N-1$}{
%\hspace{1cm}$\alpha[i]=R[i,i+1]/R[i,i]-R[i-1,i]/R[i-1,i-1];$\\
%\hspace{1cm}$\beta[i]=(R[i+1,i+1]/R[i,i])^2;$ }
%$\alpha[N]=R[N,N+1]/R[N,N]-R[N-1,N]/R[N-1,N-1];$
%\end{algorithm}
%Based on $2N$ moments of variable $X$, Algorithm~\ref{alg:1} is very
%easy to implement. It relies only on the Cholesky decomposition
%which is a standard routine. A faster implementation for the
%standard normal distribution used here is to use the corresponding
%known values $\alpha_i=0$ and $\beta_i=i$ or to use the tabulated
%values of the collocation points directly as given in
%Appendix~\ref{appendix:CollPoints}.
%
The recurrence coefficients $\alpha_i$ and $\beta_i$ are often called the
Darboux coefficients. The choice of these coefficients guarantees that
the generated polynomials ${p}_n(\vartheta)$ are orthogonal. 

%The proof of this
%is by induction~\citep{Forsythe1957}.

%The iteration in~(\ref{psi:reccurence}) may be expressed in matrix-vector
%notation as,
%\begin{equation}
%\label{eqn:Psi_Zeros} X{{\bff p}}(X)=G{{\bff p}}(X)+{p}_N(X)\bff v_N,
%\end{equation}
%where $G$ is a tridiagonal matrix and $\bff v_N=(0,0,\dots,0,1)^\T.$
%
%Matrix $G$ is symmetric when the
%polynomials ${p}_n(X)$ are orthogonal. When $G$ is not
%symmetric, we may perform a diagonal similarity transformation, which
%will yield a tridiagonal Jacobian matrix $J$, of the following
%form,
%\begin{eqnarray}
%\label{eqn:Jacobian}
%J:=P^{-1}GP=\left(\begin{array}{ccccc}\alpha_1&\beta_1&0&0&0\\
%\beta_1&\alpha_2&\beta_2&0&0\\
%0&\beta_2&\alpha_3&\beta_3&0\\
% &\ddots&\ddots&\ddots& \\
%0&0&\beta_{N-2}&\alpha_{N-1}&\beta_{N-1}\\
%0&0&0&\beta_{N-1}&\alpha_{N}\\
%\end{array}\right).
%\end{eqnarray}
The next step is to relate the coefficients $\alpha_i$ and $\beta_i$ to the zeros of the orthogonal polynomials
${p}_n(\vartheta)$, $n=0,\dots,N$, utilizing the so-called the {\em eigenvalue method}, presented in the theorem below.
\begin{thm}[Eigenvalue method]
\label{thm:eigenvalueMethod}
The zeros ${\theta_n}$, $n=1,\dots,N$, of the orthogonal polynomial
${p}_N(\vartheta)$ are the eigenvalues of the symmetric tridiagonal matrix,
\begin{eqnarray}
\label{eqn:EigenVal}
\widehat{J}:=\left(\begin{array}{ccccc}\alpha_1&\sqrt{\beta_1}&0&0&0\\
\sqrt{\beta_1}&\alpha_2&\sqrt{\beta_2}&0&0\\
%0&\sqrt{\beta_2}&\alpha_3&\sqrt{\beta_3}&0\\
 &\ddots&\ddots&\ddots& \\
0&0&\sqrt{\beta_{N-2}}&\alpha_{N-1}&\sqrt{\beta_{N-1}}\\
0&0&0&\sqrt{\beta_{N-1}}&\alpha_{N}\\
\end{array}\right),
\end{eqnarray}
i.e., $\bff \theta=(\theta_1,\dots,\theta_{N})^\T$ is a vector of eigenvalues
satisfying $\widehat{J}{\bf v}=\theta_i{\bf v}$ with $i=1,\dots,N$ for
any real vector ${\bf v}$, with  $\alpha_i$ and $\beta_i$ being the
coefficients of the three-term recurrence
relation~(\ref{psi:reccurence}).
%\begin{eqnarray*}
%{p}_{i+1}(X)=(X-\alpha_i){p}_i(X)-\beta_i{p}_{i-1}(X),\;\;\;i=2,\dots,N-1.
%\end{eqnarray*}
\begin{proof}The proof can be found in~\citep{GoWe}.
\end{proof}
\end{thm}
Based on the coefficients $\alpha_i$ and $\beta_i$, {\em the collocation points $\theta_n$ are the eigenvalues of matrix (\ref{eqn:EigenVal})}. Eigenvalue calculation for a tridiagonal matrix in Theorem~\ref{thm:eigenvalueMethod} is performed by, e.g. the Lanczos algorithm.

The final step is to determine the corresponding weights, $\omega_n$. This can be done based on the Theorem~\ref{thm:weights} below,
\begin{thm}[Optimal weights $\omega_n$]
\label{thm:weights}
Let ${\bf v}^{(n)}:=[v_1^{(n)},\dots,v_N^{(i)}]^\T$ be an eigenvector of $\widehat J$ in ~(\ref{eqn:EigenVal}) for the eigenvalue $\theta_n$,  $\widehat{J}{\bf v}^{(n)}=\theta_n{\bf v}^{(n)}$. Then, the weights $\omega_n$ are given by: $\omega_n = ({\bf v}^{(n)}_1)^2,\;\;\;n=1,\dots,N.$
\begin{proof}
Proof can be found in~\cite{bulirsch2002introduction} (p.179).
\end{proof}
\end{thm}
The pseudo algorithm to compute the pairs $\{\omega_n,\theta_n\}_{n=1}^N$ is provided in Algorithm~\ref{alg:collPoints} below.

%\begin{algorithm}
%\caption{Calculation of the collocation points ${\bff x}$ for given vectors $\bff \alpha$ and $\bff \beta$} \label{alg:2} \small {\text
%// Construction of the matrix $\widehat{J}$:}\\
%$J     = zeros(N,N), J[1,1]= \alpha[1], J[1,2]=\sqrt{\beta(1)}$\\
%\For{$i=2\ldots N-1$}{
%     $J[i,i-1]=\sqrt{\beta[i-1]}, J[i,i]=\alpha[i], J[i,i+1]=\sqrt{\beta[i]}$}
%$J[N,N]=\alpha[N], J[N,N-1]=\sqrt{\beta[N-1]};$
%\\ // Find the collocation points ${\bff x}$\\
%\KwResult{${\bff x}=eig(J)$}
%\end{algorithm}

%In Appendix~\ref{app:calculationOfAlphaAndBeta} an algorithm for computation of $\alpha$ and $\beta$ is presented. Moreover.

\begin{algorithm}
\caption{Calculation of optimal quadrature weights and nodes $\{\omega_n,\theta_n\}_{n=1}^N$, The accompanying Python and MATLAB codes can be found at~\url{https://github.com/LechGrzelak/Randomization}.} \small {\text
// Construction of the matrix $M$ with size $(N+1\times N+1)$ :}\\
\For{$i=1\ldots N+1$}
{
\For{$j=1\ldots N+1$}
{
 $M[i,j]=\E[\vartheta^{i+j-2}]$ 
 }
}
 // Calculate upper diagonal Cholesky matrix from $M$:\\
\KwResult{$M=R^\T R$} // Find $\bff \alpha$ and $\bff \beta$ from matrix
$R$, according to Equation~(\ref{eqn:alphaBeta})\\
$\alpha[1]=R[1,2]/R[1,1]$\\
$\beta[1]=(R[2,2]/R[1,1])^2$\\
\For{$i=2:N-1$}{
\hspace{1cm}$\alpha[i]=R[i,i+1]/R[i,i]-R[i-1,i]/R[i-1,i-1];$\\
\hspace{1cm}$\beta[i]=(R[i+1,i+1]/R[i,i])^2;$ }
$\alpha[N]=R[N,N+1]/R[N,N]-R[N-1,N]/R[N-1,N-1];$\\
{\text // Construction of the matrix $\widehat{J}$, according to Equation~(\ref{eqn:EigenVal}):}\\
$J     = zeros(N,N), J[1,1]= \alpha[1], J[1,2]=\sqrt{\beta(1)}$\\
\For{$i=2\ldots N-1$}{
     $J[i,i-1]=\sqrt{\beta[i-1]}, J[i,i]=\alpha[i], J[i,i+1]=\sqrt{\beta[i]}$}
$J[N,N]=\alpha[N], J[N,N-1]=\sqrt{\beta[N-1]};$
\\ // Find the collocation points ${\theta}_n$ and the corresponding weights ${\omega}_n$\\
${\bf \theta}$ is computed from ${\bf x}=EigenValues(J)$\\
${\bf \omega}$ is obtained by squaring the first row from eigenvalues vector\\
\KwResult{$\{\omega_n,\theta_n\}_{n=1}^N$}
\label{alg:collPoints}
\end{algorithm}

%%%%%%%%%%%%%%%%%%%%%%%%%%%%%%%%%%%%%%%%%%%%%%%%%%%%%%%%%%%%%%%%%%%%%%%%%%%%%
\subsection{\bf Impact of randomizers on implied volatilities}
\label{app:randomizersImpact}
\begin{figure}[h!]
  \centering
      \includegraphics[width=0.45\textwidth]{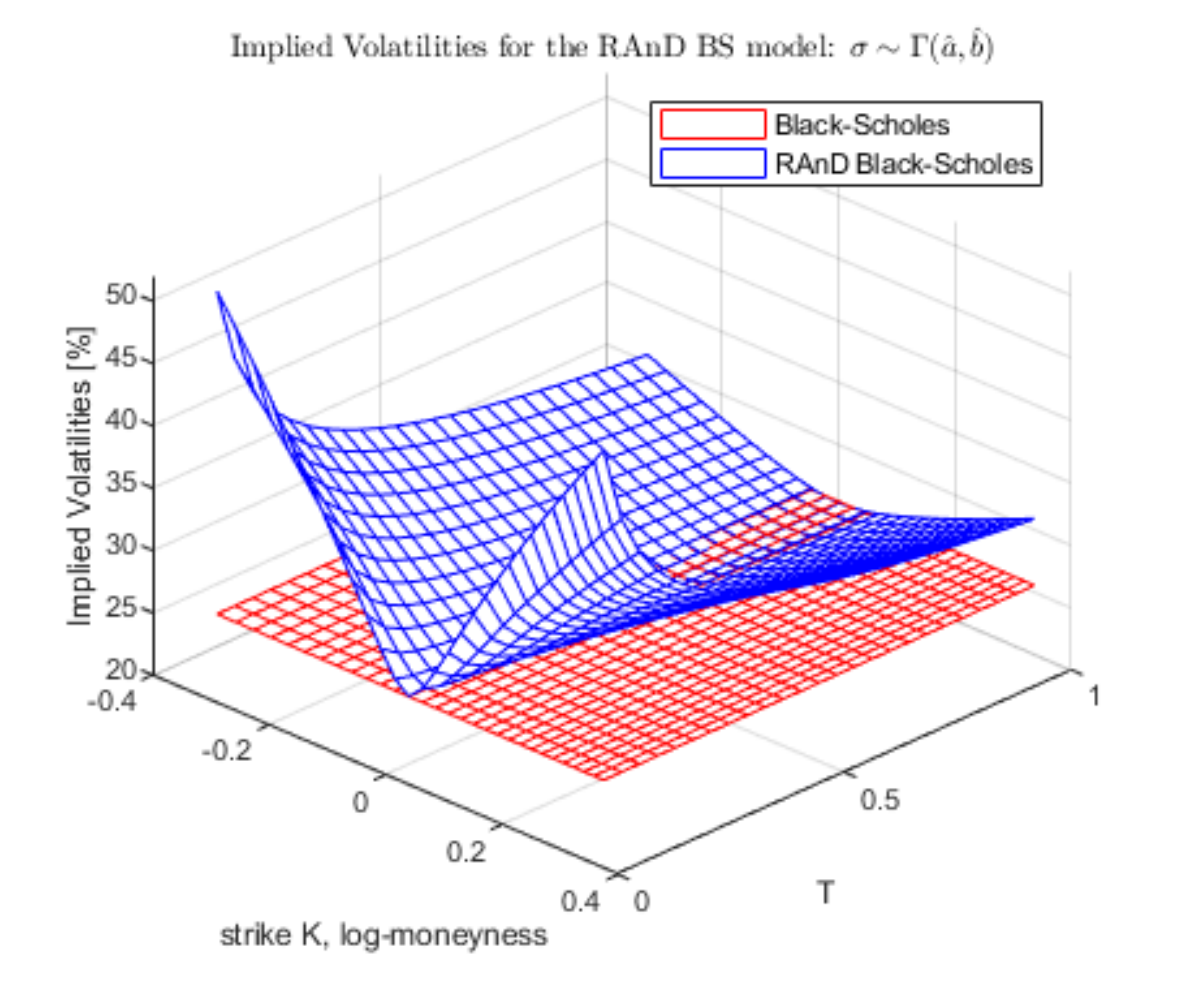}
      \includegraphics[width=0.45\textwidth]{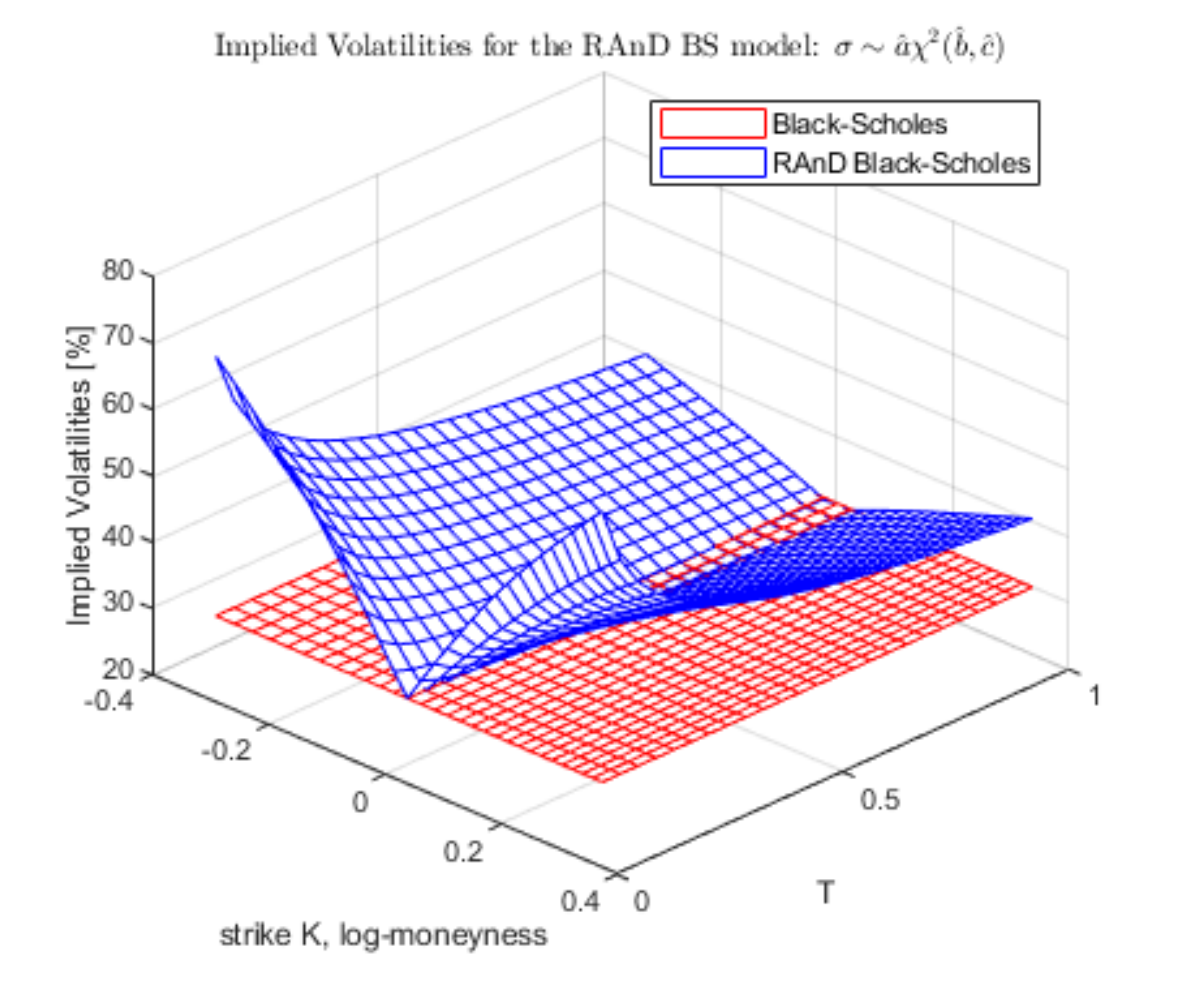}
          \caption{Implied volatility surface for the RAnD BS model for different randomizers of $\sigma.$}
      \label{fig:RandBS_2}
\end{figure}

%%%%%%%%%%%%%%%%%%%%%%%%%%%%%%%%%%%%%%%%%%%%%%%%%%%%%%%%%%%%%%%%%%%%%%%%%%%%%
In this section, we illustrate the impact of randomizers on the implied volatilities under the Black-Scholes model. In particular we consider $\sigma\sim~\Gamma(\hat a,\hat b)$ and $\sigma\sim \hat a\chi^2(\hat b,\hat c)$. The results are reported in Figure~\ref{fig:RandBS_2}.

\subsection{\bf ChF for the Bates model}
%%%%%%%%%%%%%%%%%%%%%%%%%%%%%%%%%%%%%%%%%%%%%%%%%%%%%%
For the Bates model, the complex valued functions $\bar C(u,\tau)$, $\bar A_H(u,\tau)$ are known explicitly and are given by:
\label{sec:AC}
\begin{eqnarray*}
	\bar C(u,\tau)&=&\frac{1-\e^{- D \tau}}{\gamma^2(1-g\e^{- D \tau})}\left(\kappa-\gamma\rho iu- D \right),\\
	\bar A_H(u,\tau)&=&r(iu-1)\tau+\frac{\kappa\bar{v}\tau}{\gamma^2}\left(\kappa-\gamma\rho iu- D \right)-\frac{2\kappa\bar{v}}{\gamma^2}\log\left(\frac{1-g\e^{- D \tau}}{1-g}\right),\\
	\bar A_B(u,\tau)&=&\bar A_H({u},\tau)-\lambda iu\tau \left(\e^{\mu_J+\frac12\sigma_J^2}-1\right)+\lambda\tau\left(\e^{iu\mu_J-\frac12\sigma_J^2u^2}-1\right). \end{eqnarray*} 
for $ D =\sqrt{(\kappa-\gamma\rho iu)^2+(u^2+iu)\gamma^2}$ and
$g=\displaystyle\frac{\kappa-\gamma\rho iu- D }{\kappa-\gamma\rho iu+ D }$ and where $\tau=T-t.$

%%%%%%%%%%%%%%%%%%%%%%%%%%%%%%%%%%%%%%%%%%%%%%%%%%%%%%%%%%%%%%%%%%
\subsection{\bf Proof of Lemma~\ref{lem:vix}}
\label{appendix:vix}
%%%%%%%%%%%%%%%%%%%%%%%%%%%%%%%%%%%%%%%%%%%%%%%%%%%%%%%%%%%%%%%%%%
Using the dynamics of the Bates model in Equation~(\ref{eqn:Bates}) and the definition of the VIX~(\ref{eqn:vix_def}) we have:
\begin{eqnarray}
\log \frac{S(T)}{S(t)}=-\lambda\E[\e^J-1](T-t)-\frac12\int_t^Tv(u)\du + \int_t^T\sqrt{v(u)}\dW(u)+\int_t^TJ\d X_{\mathcal{P}}(u),
\end{eqnarray}
with $\E_t[\e^J]=\e^{\mu_J+\frac12\sigma_J^2}$ and all the model parameters as specified in~(\ref{eqn:Bates}). By taking the expectations, we find,
\begin{eqnarray}
\label{eqn:CIRIntegration}
\E_t\left[\log \frac{S(T)}{S(t)}\right]=-\lambda\left(\e^{\mu_J+\frac12\sigma_J^2}-\mu_J-1\right)(T-t)-\frac12\int_t^T\E_t[v(u)]\du.
\end{eqnarray}
The expectation, conditioned on time $t$, of the CIR process is known analytically and it is given by:
\[\E_t[v(T)]=\bar c(t,T)(\delta+\bar \kappa(t,T)),\]
\begin{equation}
\bar c(t,T)=\frac{1}{4\kappa}\gamma^2(1-\e^{-\kappa (t-T)}),\;\;
	\delta=\frac{4\kappa\bar{v}}{\gamma^2},\;\; \bar \kappa(t,T)=\frac{4\kappa v(t)
\e^{-\kappa (T-t)}}{\gamma^2(1-\e^{-\kappa (T-t)})}.
\end{equation}
By integration over the expectation in Equation~(\ref{eqn:CIRIntegration}) and substitution to  definition of VIX in~(\ref{eqn:vix_def}) we find:
\begin{eqnarray}
{{\vix}}^2(t,T)=100^2\times\frac{-2}{T-t}\E_t\left[\log \frac{S(T)}{S(t)}\right]=a(t,T)v(t)+b(t,T)+c,
\end{eqnarray}
%\[\vix^2(t,T)=a(t,T)v(t)+b(t,T)+c,\]
with \[a(t,T)=\frac{1-\e^{-\kappa(T-t)}}{\kappa(T-t)},\;\;\;b(t,T)=\bar{v}\left(1-a(t,T)\right),\;\;c=2\lambda\left(\e^{\mu_J+\frac12\sigma_J^2}-\mu_J-1\right).\]

Since the CIR process $v(t)|v(t_0), \; t>0$, follows a scaled, with $\bar c(t_0,t)$, non-central chi-square distribution, $\chi^2(\delta,\bar \kappa(t_0,t))$, where $\delta$ is the ``degrees of freedom'' parameter and the noncentrality parameter is
$\bar \kappa(t_0,t)$, i.e., 
\begin{equation}
\label{chapter3:sigma_chi2a}
\begin{aligned}
v(t)|v(t_0)\sim \bar c(t_0,t)\chi^2\left(\delta,\bar \kappa(t_0,t)\right),&&t>t_0.
\end{aligned}
\end{equation}
Therefore the CDF for $\VIX$ is given by:
\begin{eqnarray*}
\P\left[\vix(t,T)\leq x\right]&=&\P\left[\sqrt{a(t,T)v(t)+b(t,T)+c}\leq x\right]\\&=&\P\left[v(t)\leq \frac{1}{a(t,T)}\left(x^2-b(t,T)-c\right)\right]=F_{v(t)}\left(\frac{1}{a(t,T)}\left(x^2-b(t,T)-c\right)\right).
\end{eqnarray*}
Since the unconditional CDF for $v(t)$ is known explicitly, we have
\begin{equation}
\label{chapter3:eqn:20}
F_{v(t)}(x)=\Q[v(t)\leq
	x]=\Q\left[\chi^2\left(\delta,\bar \kappa(t_0,t)\right)\leq
	\frac{x}{\bar c(t_0,t)}\right]=F_{\chi^2(\delta,\bar \kappa(t_0,t))}\left(\frac{x}{\bar c(t_0,t)}\right),
\end{equation}
thus finally we have:
\begin{eqnarray*}
F_{\vix}(x)=F_{\chi^2(\delta,\bar \kappa(t_0,t))}\left(\frac{x^2-b(t,T)-c}{a(t,T)\bar c(t_0,t)}\right).
\end{eqnarray*}
By differentiation, the probability density function reads:
\begin{eqnarray*}
	f_{\vix}(x)&:=&\frac{\d}{\d x}F_{\chi^2(\delta,\bar \kappa(t_0,t))}\left(\frac{x^2-b(t,T)-c}{a(t,T)\bar c(t_0,t)}\right)\\
	&=&\frac{2x}{a(t,T)\bar c(t_0,t)}f_{\chi^2(\delta,\bar \kappa(t_0,t))}\left(\frac{x^2-b(t,T)-c}{a(t,T)\bar c(t_0,t)}\right)=:	2\alpha_1xf_{\chi^2(\delta,\bar \kappa(t_0,t))}\left(\alpha_1(x^2-\alpha_2)\right),
	%\frac{x}{\alpha_1}f_{\chi^2(\delta,\bar \kappa(t_0,t))}\left(\frac{x^2-\alpha_2}{\alpha_3}\right)
\end{eqnarray*}
with $\alpha_1=\frac{1}{a(t,T)\bar{c}(t_0,t)}$ and $\alpha_2=b(t,T)+c$.

Having all the ingredients at hand we derive the ChF for $\vix^2(t,T)$. By definition we find:
\begin{eqnarray*}
\phi_{\vix^2(t,T)}(u)&=&\E_{t_0}\left[\e^{iu(a(t,T)v(t)+b(t,T)+c)}\right]\\
&=&\e^{iu(b(t,T)+c)}\E_{t_0}\left[\e^{iua(t,T)v(t)}\right]=\e^{iu(b(t,T)+c)}\mathcal{M}_{v(t)}(iua(t,T)),
\end{eqnarray*}
where $\mathcal{M}_{v(t)}(iua(t,T))$ is the moment-generating function of $v(t)$, known explicitly (see~\cite{OosterleeGrzelakBook} p.314):
\[\mathcal{M}_{v(t)}(u)=\E_{t_0}\left[\e^{u v(t)}\right]=\left(\frac{1}{1-2u\bar{c}(t_0,t)}\right)^{\frac12\delta}
\exp\left(\frac{u\bar{c}(t_0,t)\bar{\kappa}(t_0,t)}{1-2u\bar{c}(t_0,t)}\right).\]

%%%%%%%%%%%%%%%%%%%%%%%%%%%%%%%%%%%%%%%%%%%%%%%%%%%
%\section{ChF for the randomized Black-Scholes model}
%%%%%%%%%%%%%%%%%%%%%%%%%%%%%%%%%%%%%%%%%%%%%%%%%%%
%The randomized ChF is given by:
%\begin{eqnarray*}
%\phi_{{X}}({ u};t,T)&=&\sum_{n=1}^{N}\omega_n\phi_{{ X}|\sigma=\sigma_n}({ %u};t,T)+\epsilon_N,
%\end{eqnarray*}
%The conditional ChF form the $\log S(t)$ under the Black-Scholes model reads:
%\begin{eqnarray*}
%\phi_{X|\sigma=\sigma_n}(u;t,T) = %\exp\big({iuX(t)+(r-\frac12\sigma_n^2)iu(T-t)-\frac12\bar\sigma^2u^2(T-t)}\big)
%\end{eqnarray*}
%so the randomized ChF reads now:
%\begin{eqnarray*}
%\phi_{{X}}({ u};t,T)&=&\sum_{n=1}^{N}\exp\big(\log %\omega_n+{iuX(t)+(r-\frac12\sigma_n^2)iu(T-t)-\frac12\bar\sigma^2u^2(T-t)}\big)+\%epsilon_N,
%\end{eqnarray*}

%%%%%%%%%%%%%%%%%%%%%%%%%%%%%%%%%%%%%%%%%%%%%%%%%%%%%%
\subsection{\bf Implied volatilities for RAnD BS model with uniform randomizer}
\label{sec:RAnDBS_uniform}
%%%%%%%%%%%%%%%%%%%%%%%%%%%%%%%%%%%%%%%%%%%%%%%%%%%%%%%

\begin{table}[htb!]
\centering\footnotesize
\caption{\footnotesize  Maximum implied volatility, $\text{IV}(\cdot)$, error computed against Monte Carlo results with 10 million samples, with $K=S_0\e^{0.1\sqrt{T}\delta}$, $\delta=[-3,-2,-1,-0.5,0,0.5,1,2,3]$.}
\begin{tabular}{c|c|c|c|c|c|c|c|c}
\multicolumn{8}{c}{$\text{error} =\max_i{\big|\text{IV}(t_0,K_i)-\text{IV}_{\text{ref}}(t_0,K_i)\big|}$}\\
$\sigma\sim\mathcal{U}([a,b])$&$N=2$&$N=3$&$N=4$&$N=5$&$N=6$&$N=7$&$N=8$&$N=9$\\\hline\hline
$T=1d$&	$0.16\; \%$&$0.04\; \%$&$0.02\; \%$&$0.01\; \%$&$0.01\; \%$&$0.01\; \%$& $0.01\; \%$& $0.01\; \%$\\
$T=1w$& $0.16\; \%$&$0.04\; \%$&$0.01\; \%$&$0.01\; \%$&$0.00\; \%$&$0.00\; \%$& $0.00\; \%$& $0.00\; \%$\\
$T=2w$& $0.16\; \%$&$0.03\; \%$&$0.02\; \%$&$0.02\; \%$&$0.02\; \%$&$0.02\; \%$& $0.02\; \%$& $0.02\; \%$\\
$T=1m$& $0.15\; \%$&$0.03\; \%$&$0.01\; \%$&$0.01\; \%$&$0.00\; \%$&$0.01\; \%$& $0.01\; \%$& $0.01\; \%$\\
$T=3m$& $0.15\; \%$&$0.03\; \%$&$0.01\; \%$&$0.01\; \%$&$0.01\; \%$&$0.01\; \%$& $0.01\; \%$& $0.01\; \%$\\
$T=6m$& $0.15\; \%$&$0.03\; \%$&$0.01\; \%$&$0.00\; \%$&$0.00\; \%$&$0.00\; \%$& $0.00\; \%$& $0.00\; \%$\\
$T=12m$& $0.15\; \%$&$0.03\; \%$&$0.01\; \%$&$0.01\; \%$&$0.01\; \%$&$0.01\; \%$& $0.01\; \%$& $0.01\; \%$\\
\end{tabular}
\label{Tab:RAnd_ImpliedVolsUniform}
\end{table}

%%%%%%%%%%%%%%%%%%%%%%%%%%%%%%%%%%%%%%%%%%%%%%%%%%%%%%%%%%%%%%%%%%%%%%
\subsection{\bf Impact of randomization on Bates model parameters}
\label{sec:appendix_impact}
\begin{figure}[h!]
  \centering
    \includegraphics[width=0.45\textwidth]{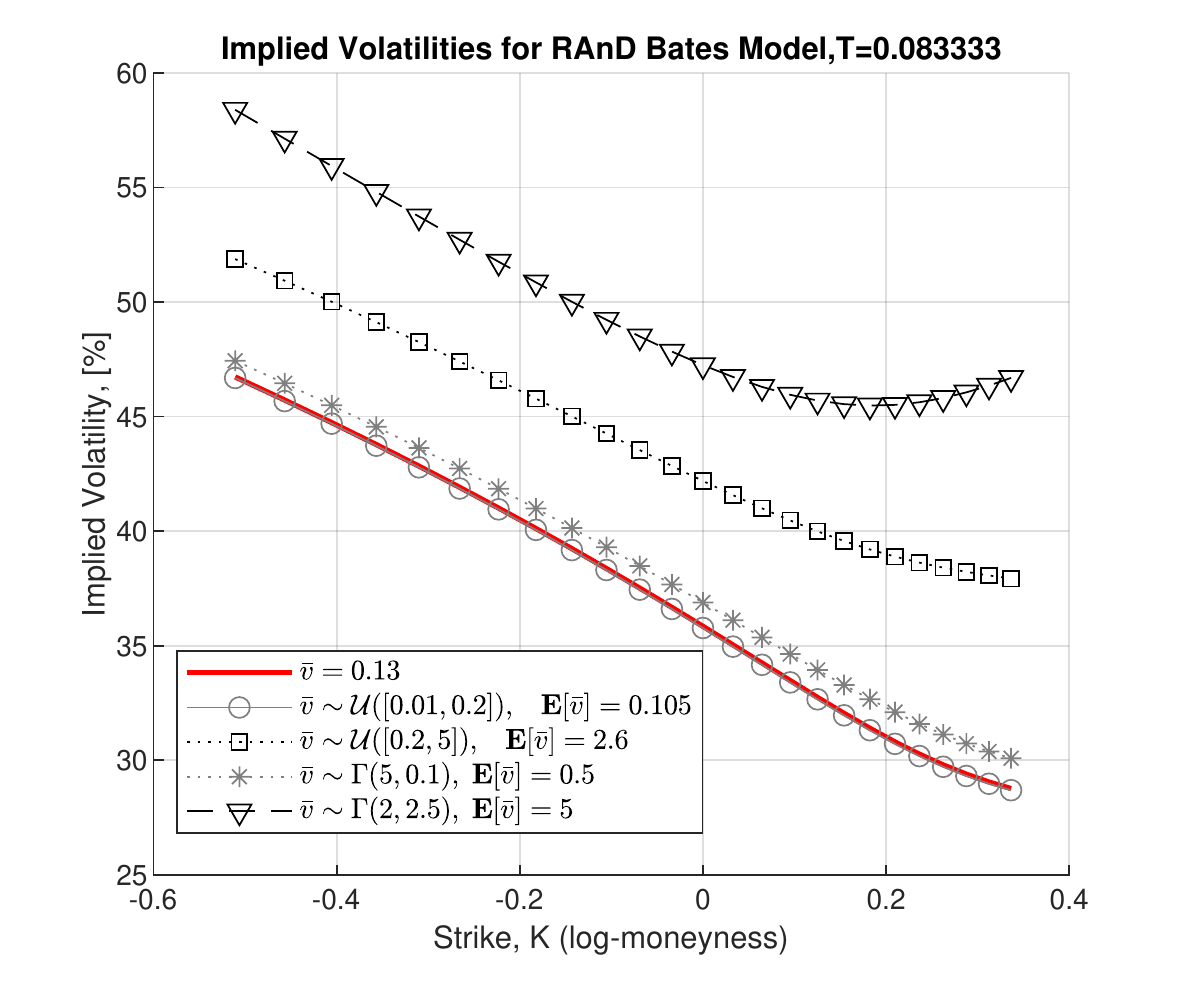}
    \includegraphics[width=0.45\textwidth]{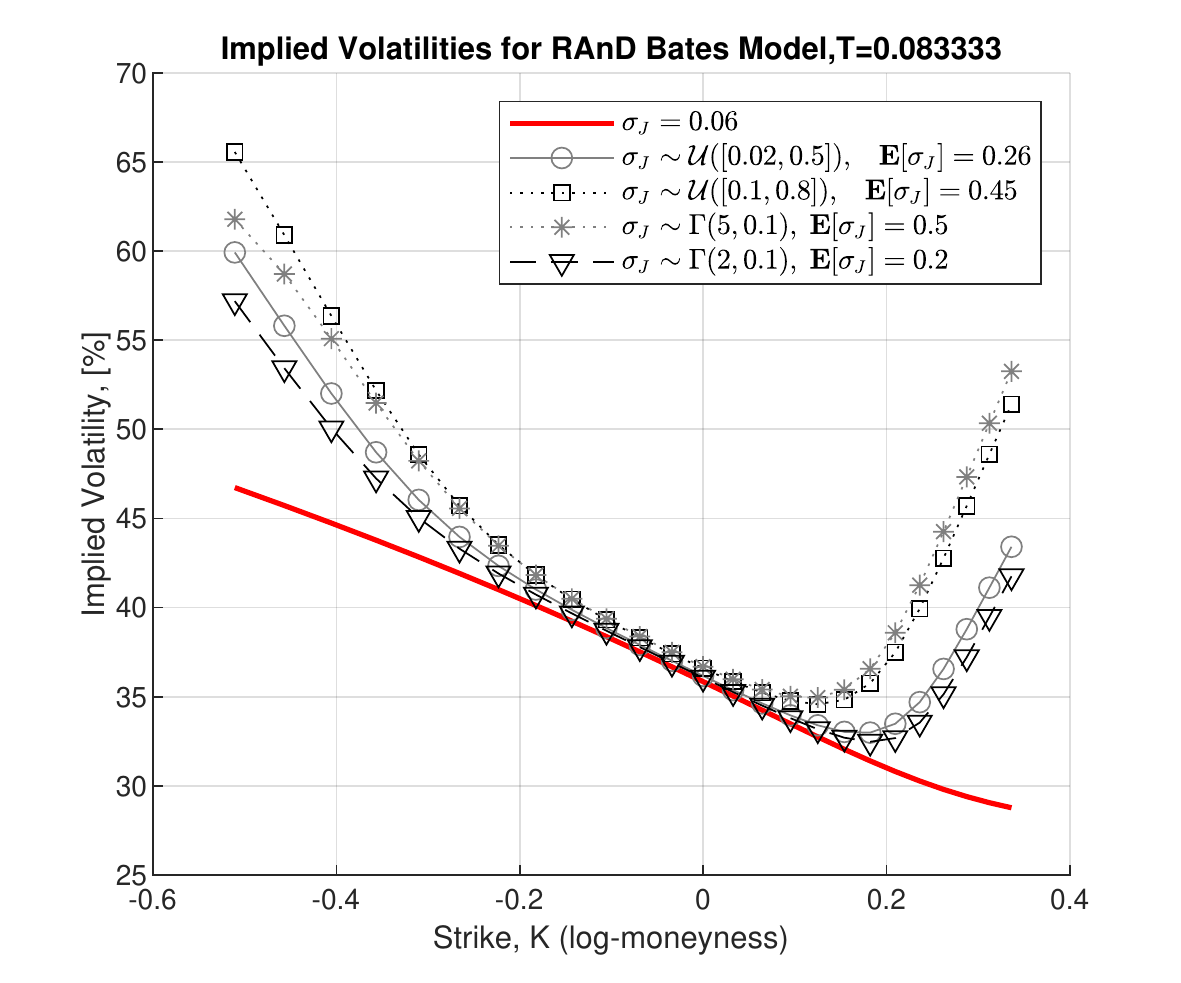}
      \caption{Impact of randomized parameters on implied volatlities. Left: randomized {\it long-term vol}, $\bar{v}$. Right: randomized {\it jump's size vol}, $\sigma_J$.}
      \label{fig:impact3}
\end{figure}

\begin{figure}[h!]
  \centering
    \includegraphics[width=0.45\textwidth]{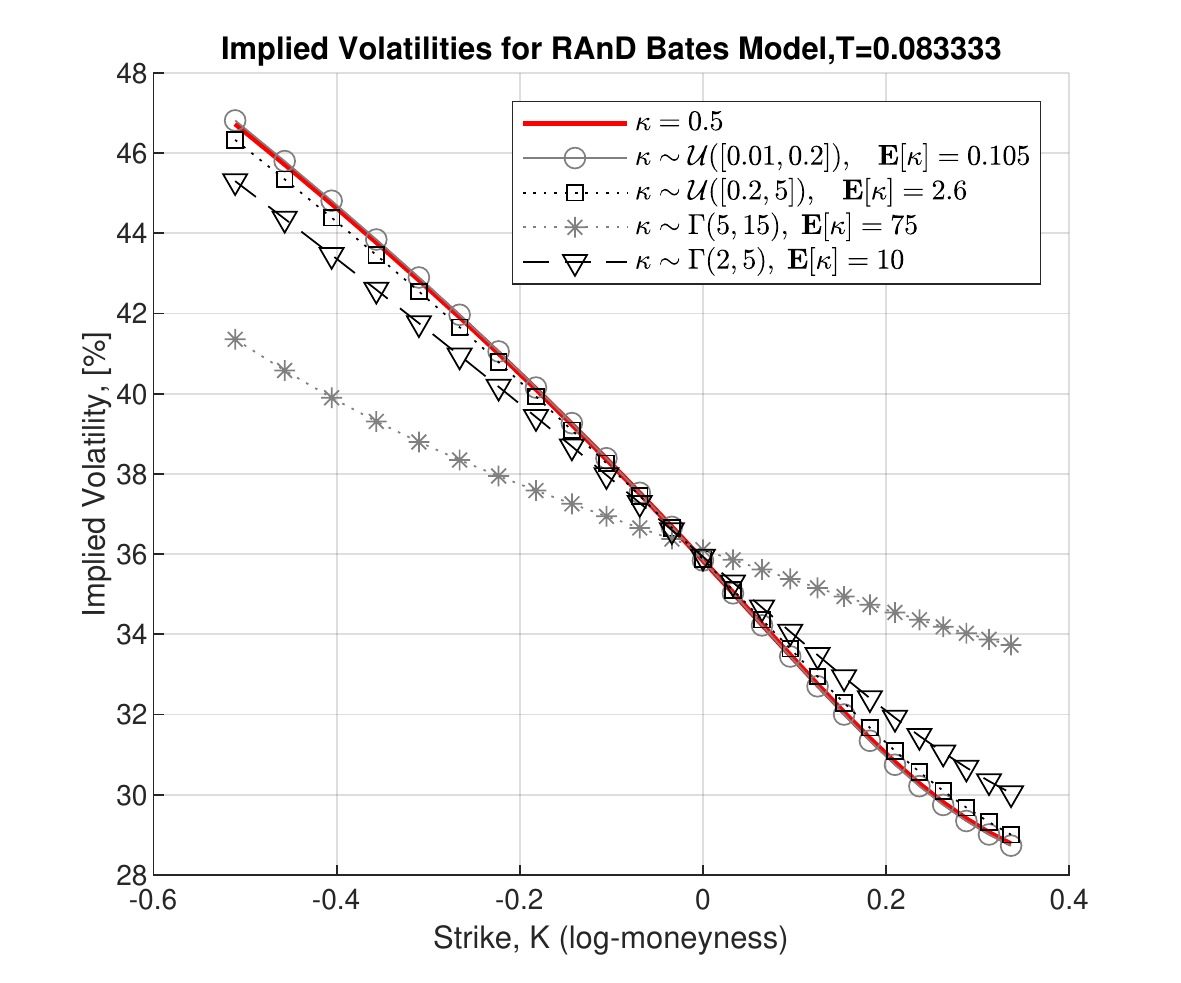}
    \includegraphics[width=0.45\textwidth]{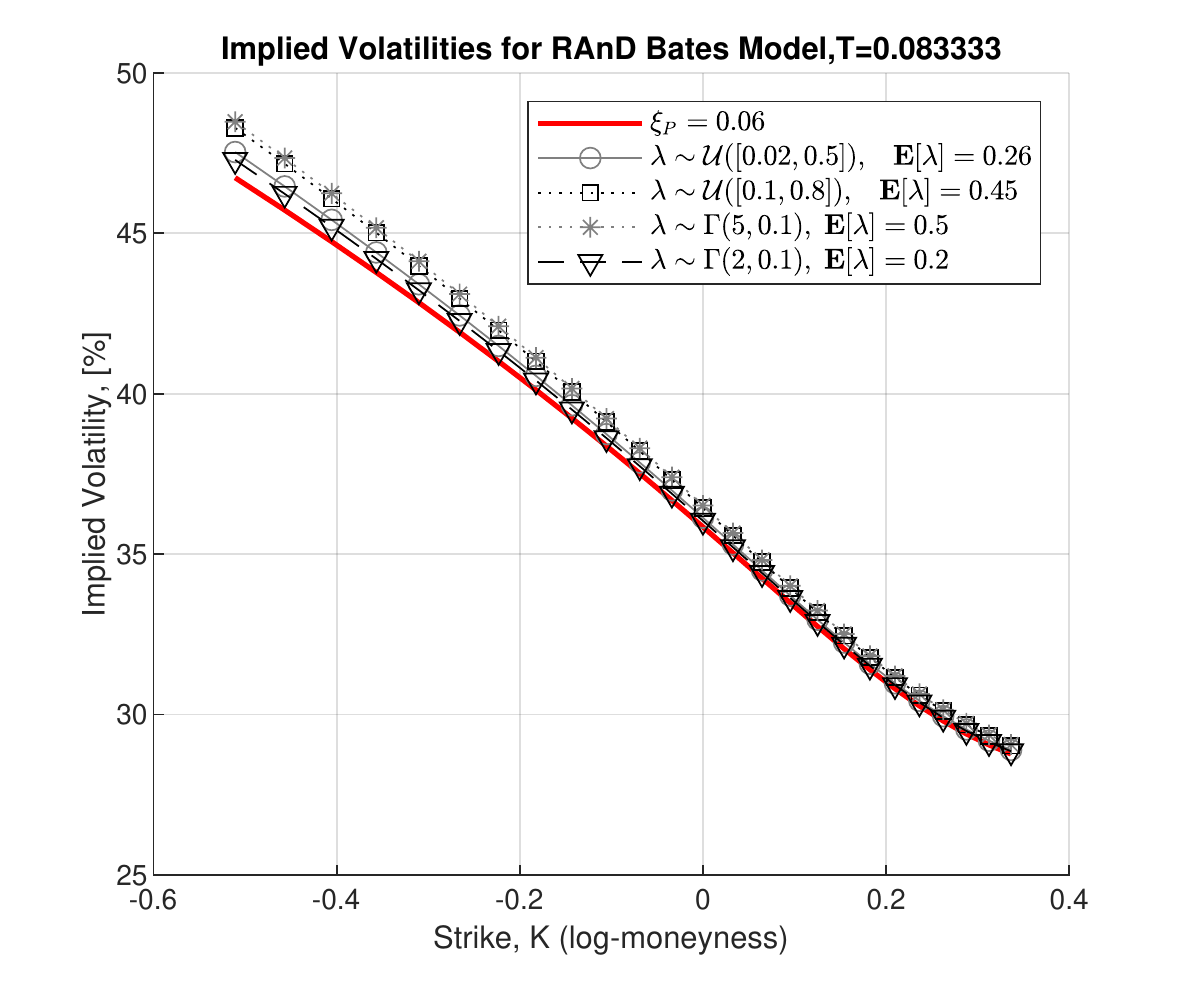}
      \caption{Impact of randomized parameters on implied volatlities. Left: randomized {\it speed of mean reversion}, $\kappa$. Right: randomized {\it jump's intensity}, $\lambda$.}
      \label{fig:impact4}
\end{figure}

%%%%%%%%%%%%%%%%%%%%%%%%%%%%%%%%%%%%%%%%%%%%%%%%%%%%%%%%%%%%%%%%%%%%%%%%%
The impact of different randomizers on implied volatilities under the Bates model is presented in Figure~\ref{fig:impact3} and  Figure~\ref{fig:impact4}.

\end{document}